\documentclass[acmsmall]{acmart}
\usepackage{indentfirst} 
\usepackage{algorithmic}
\usepackage{textcomp}
\usepackage[ruled,vlined,lined,commentsnumbered]{algorithm2e}
\usepackage{booktabs}
\usepackage{moreverb}
\usepackage{fontenc}

\usepackage{amsmath,amssymb,amsfonts}
\usepackage{fancybox}
\usepackage{wrapfig}
\usepackage{color}
\usepackage{xcolor,colortbl}
\usepackage{array}
\usepackage{multirow}
\usepackage{multicol}
\usepackage{makecell}
\usepackage{graphicx}
\usepackage{rotating}
\usepackage{setspace}
\usepackage{soul}
\usepackage{diagbox}
\usepackage{balance}
\usepackage{csvsimple}
\usepackage{longtable}
\usepackage[skip=0pt,labelfont=bf]{caption}
\usepackage{url}
\usepackage{lscape}
\usepackage{rotating}
\usepackage{tikz}
\usepackage[shortlabels]{enumitem}
\usepackage{subcaption}
\usepackage{listings}
\usepackage[most]{tcolorbox}
\usepackage{threeparttable}
\usepackage{bclogo}
\usepackage{marvosym}
\usepackage{pifont}
\usepackage{listings}
\usepackage{siunitx}
\usepackage{subcaption}

\newcommand*{\ie}{i.e., }

\newcommand{\toolname}{\textbf{\texttt{PacVD}}\xspace}

\newcommand{\etal}{\emph{et~al.}\xspace}

\newcommand{\intuition}[1]{
\begin{tcolorbox}[tile,size=fbox,boxsep=1.5mm,boxrule=0pt,top=0pt,bottom=0pt,
borderline west={1mm}{-2pt}{black!50!white},colback=black!5!white]
\em #1
\end{tcolorbox}
}

\definecolor{codegreen}{rgb}{0,0.6,0}
\definecolor{codegray}{rgb}{0.5,0.5,0.5}
\definecolor{codepurple}{rgb}{0.58,0,0.82}
\definecolor{backcolour}{rgb}{0.95,0.95,0.92}
\definecolor{lightgreen}{HTML}{99d8c9}
\definecolor{lightgray}{rgb}{0.95, 0.95, 0.95}

\AtBeginDocument{%
  }

\setcopyright{acmlicensed}
\copyrightyear{2018}
\acmYear{2018}
\acmDOI{XXXXXXX.XXXXXXX}

\acmJournal{JACM}
\acmVolume{37}
\acmNumber{4}
\acmArticle{111}
\acmMonth{8}

\begin{document}

\title{Context-Enhanced Vulnerability Detection Based on Large Language Model}



\author{Yixin Yang}
\affiliation{%
  \institution{State Key Laboratory of Complex \& Critical Software Environment (CCSE), Beihang University}
  \city{Beijing 100191}
  \country{China}
}
\email{yixinyang@buaa.edu.cn}

\author{Bowen Xu}
\affiliation{%
  \institution{State Key Laboratory of Complex \& Critical Software Environment (CCSE), Beihang University}
  \city{Beijing 100191}
  \country{China}
}
\email{cheneytsu@buaa.edu.cn}



\author{Xiang Gao\textsuperscript{*}}
\affiliation{%
  \institution{State Key Laboratory of Complex \& Critical Software Environment (CCSE), Beihang University}
  \city{Beijing 100191}
  \country{China}
}
\affiliation{%
  \institution{Hangzhou Innovation Institute of Beihang University}
  \city{Hangzhou 310056}
  \country{China}
}
\email{xiang_gao@buaa.edu.cn}

\author{Hailong Sun}
\affiliation{%
  \institution{State Key Laboratory of Complex \& Critical Software Environment (CCSE), Beihang University}
  \city{Beijing 100191}
  \country{China}
}
\affiliation{%
  \institution{Hangzhou Innovation Institute of Beihang University}
  \city{Hangzhou 310056}
  \country{China}
}
\email{sunhl@buaa.edu.cn}

\thanks{*These authors are co-corresponding authors.}

\begin{abstract}
Vulnerability detection is a critical aspect of software security. 
Accurate detection is essential to prevent potential security breaches and protect software systems from malicious attacks.
Recently, vulnerability detection methods leveraging deep learning and large language models (LLMs) have garnered increasing attention. 
However, existing approaches often focus on analyzing individual files or functions, which limits their ability to gather sufficient contextual information. 
Analyzing entire repositories to gather context introduces significant noise and computational overhead. 
To address these challenges, we propose a context-enhanced vulnerability detection approach that combines program analysis with LLMs.
Specifically, we use program analysis to extract contextual information at various levels of abstraction, thereby filtering out irrelevant noise. 
The abstracted context along with source code are provided to LLM for vulnerability detection.
We investigate how different levels of contextual granularity improve LLM-based vulnerability detection performance. 
Our goal is to strike a balance between providing sufficient detail to accurately capture vulnerabilities and minimizing unnecessary complexity that could hinder model performance. 
Based on an extensive study using GPT-4, DeepSeek, and CodeLLaMA with various prompting strategies, our key findings includes: (1) incorporating abstracted context significantly enhances vulnerability detection effectiveness; (2) different models benefit from distinct levels of abstraction depending on their code understanding capabilities; and (3) capturing program behavior through program analysis for general LLM-based code analysis tasks can be a direction that requires further attention.
\end{abstract}





\maketitle

\section{Introduction}


Memory safety vulnerabilities constitute a persistent and critical class of software security flaws in computing systems. Despite extensive research and mitigation strategies, these vulnerabilities remain predominant in recent Common Vulnerabilities and Exposures (CVE) announcements~\cite{CVE, 10.1145/3460319.3464807}. 
The fundamental issue stems from memory-unsafe languages such as C and C++, which prioritize performance and low-level memory control over safety guarantees. 
These languages permit direct memory manipulation during buffer operations, allocation, and pointer management, creating significant security risks. 
Although modern programming languages incorporate memory safety features, legacy codebases and performance-sensitive applications continue to rely on memory-unsafe languages, perpetuating their vulnerability to exploitation.

Static vulnerability detection is an effective technique that identifies potential security vulnerabilities by analyzing code without executing the program.
Current approaches to static vulnerability detection can be broadly categorized into static program analysis, machine learning, and deep learning-based methods.
Traditional static program analysis methods often rely on expert-defined rules, which, while effective, are time consuming, labor intensive, and difficult to generalize across different codebases~\cite{shi2018pinpoint}.
Machine learning-based methods for vulnerability detection utilize manually defined quantitative code features (\ie cyclomatic complexity, number of nested loops, the maximum count of control/data structures, etc.) combined with traditional machine learning algorithms (\ie SVM, KNN, etc.). 
They offer better generalization capabilities~\cite{du2019leopard}.
However, due to their shallow architectures and the coarse-grained features they extract, they often fail to achieve high detection accuracy.
Deep learning-based vulnerability detection methods have recently surpassed traditional approaches, leveraging rich semantic features extracted from source code (\ie Abstract Syntax Trees (ASTs), Control Flow Graphs (CFGs), Program Dependency Graphs (PDGs), etc.) and deep learning algorithms~\cite{li2018vuldeepecker, li2022sysevr, chakraborty2021deep, zhou2019devign}.
These approaches significantly improve both the effectiveness and interpretability of vulnerability detection by capturing richer structural and semantic information.

Despite the progress made by learning-based methods, several limitations still remain.
First, function-level and file-level vulnerability detection methods often rely on using individual pre- and post-patch functions as the input to the model~\cite{li2018vuldeepecker, li2022sysevr, chakraborty2021deep, zhou2019devign}.
However, these methods lack sufficient context to capture the root cause of vulnerabilities.
In many cases, the root cause of a vulnerability might not be accurately reflected by the modified function alone, as it often involves cross-function interactions.
Additionally, these methods tend to focus heavily on understanding token-level information, which makes it challenging for the models to generalize to new vulnerabilities not seen during training, resulting in a high number of false positives and false negatives.
Second, vulnerability detection methods at the repository level aim to provide a more comprehensive context by analyzing entire codebases~\cite{wen2024vuleval, zhou2024comparison}.
While such methods can theoretically capture cross-function or cross-file interactions, these methods suffer from substantial noise.
For instance, repository-level analysis involves a lot of irrelevant information that can obscure meaningful patterns, making it difficult for models to accurately detect vulnerabilities.
Furthermore, analyzing large code repositories, especially those with millions of lines of code, presents significant challenges due to compilation issues, static analysis overhead, and the  ``path explosion'' problem in control flow analysis, which results in increased computational costs and limited scalability.

Recently, large language models (LLMs), such as GPT-4~\cite{achiam2023gpt}, LLaMA~\cite{touvron2023llama} and DeepSeek~\cite{deepseekai2025deepseekr1incentivizingreasoningcapability}, have shown remarkable code comprehension capabilities, drawing increasing attention from researchers in the field of software engineering.
Trained on massive code repositories with billions of parameters, these models have the potential to capture deeper semantic information and address the shortcomings of traditional shallow learning models.
Several work has explored applying LLMs for vulnerability detection, utilizing prompt engineering~\cite{10.1145/3639478.3643065,DBLP:journals/corr/abs-2403-17218,10.1145/3639476.3639762,10.1016/j.jss.2024.112234,Tamberg_2025} 
and model fine-tuning~\cite{du-etal-2024-generalization,Shestov2024Finetuning,Mao2024Towards,wang2024m2cvdenhancingvulnerabilitysemantic,han2024parameterefficientfinetuninglargemodels}
, and other strategies~\cite{du2024vulragenhancingllmbasedvulnerability,li2025irisllmassistedstaticanalysis}
.
However, similar to earlier deep learning models, LLMs struggle to directly infer the root causes of vulnerabilities with inadequate contextual information.
Lacking specific contextual information inspired us to investigate whether it is possible to use abstract representations of vulnerability-related contexts to enhance the reasoning capabilities of large language models.

To address this challenge, we first utilize program analysis to design a method to extract contextual information for Inter-procedural vulnerabilities.
Specifically, we conduct control flow analysis and data flow analysis on the to-be-patched functions and their callee functions within a specific iteration layer, then we abstract function calls based on primitive APIs to provide contextual information to LLMs. 
Primitive APIs, e.g., \textit{malloc}, are basic operations that commonly occur in various scenarios
The motivation for considering primitive APIs is rooted in their fundamental nature~—~ memory safety vulnerabilities often arise in the improper usage of these APIs. 
By focusing on primitive APIs, we can effectively abstract away the noise introduced by complex or domain-specific function calls, providing clearer contextual information for the LLMs to understand potential vulnerabilities.
Second, we conduct an empirical study to evaluate how different levels of primitive API abstraction and different prompt engineering strategies impact the detection capabilities of LLMs. 
The motivation for using different API abstraction level is to determine the optimal level of contextual granularity that improves vulnerability detection while avoiding excessive noise.
By evaluating various abstraction levels, we aim to identify the balance between providing sufficient detail to capture vulnerabilities accurately and reducing unnecessary complexity that may hinder model performance.
The motivation for evaluating different prompt engineering strategies is to explore how effectively tailored prompts can enhance the reasoning capabilities of LLMs for vulnerability detection. Specifically, different prompt strategies, such as Chain-of-Thought~\cite{wei2022chain}, Few-Shot Learning~\cite{brown2020language}, and In-Context Learning~\cite{sun2023does}, provide different ways to guide the model's focus and inference process. By systematically evaluating these strategies, we aim to determine the most suitable prompt techniques for different abstraction levels, maximizing the models' ability to understand complex contexts and accurately detect vulnerabilities.
Our empirical study reveals the following major findings:
\begin{enumerate}[leftmargin=*]
    \item \textbf{Primitive API Abstraction is Effective:} Using primitive API abstractions significantly enhances vulnerability detection performance across different LLMs, showing improvements in accuracy and reducing false positives compared to models without API abstractions.
    \item \textbf{Different Vulnerabilities Types Require Varying API Abstraction Levels:} Resource management-related vulnerabilities necessitate high-level abstraction, while both high- and low-level abstractions demonstrate good performance for boundary-related memory errors.
    \item \textbf{Different Models Require Different Levels of Abstraction:} For larger models like GPT-4o and DeepSeek-r1, higher-level abstractions are more suitable, while models like CodeLLaMA benefit from more detailed abstractions, as they are specifically optimized for code understanding and generation.
    \item \textbf{Different Prompt Engineering Suits for Different API Abstraction Levels:} Higher API abstraction levels, which enhance contextual information richness, favor more advanced reasoning-oriented prompt strategies.
\end{enumerate}
Based on our findings, we implement a tool named \toolname, that is, \textbf{P}rimitive \textbf{A}PI Abstraction and \textbf{C}ontext-Enhanced \textbf{V}ulnerability \textbf{D}etection method utilizing Large Language Models.
Experimental results show that our method significantly outperforms the baselines. 
Using the Chain-of-Thought prompt strategy with DeepSeek-R1, our approach achieves improvements of up to 12.77\% in accuracy, 10.05\% in precision, 9.25\% in F1 score compared to baseline methods.
Similar improvements are observed with ChatGPT-4o and CodeLLaMA, demonstrating the effectiveness of our API abstraction approach across different model architectures. 
These results suggest that combining appropriate API abstractions with well-designed prompt strategies can substantially enhance the vulnerability detection capabilities of large language models.

Our contributions can be summarized as follows:

\begin{itemize}[leftmargin=*]
 \item \textbf{Vulnerability Contextual Information Extraction Method}: We propose a novel method for extracting contextual information of to-be-patched functions, utilizing control-flow and data-flow summaries of primitive APIs, which helps in capturing the root cause of vulnerabilities.

  \item \textbf{Empirical Study:} We conduct an empirical evaluation to examine how different types of primitive API abstractions and different prompt engineering strategies affect vulnerability detection performance in large language models.

  \item \textbf{Context-Enhanced Vulnerability Detection Model:} We introduce a new context-enhanced vulnerability detection approach that utilizes primitive API abstractions to improve the reasoning and detection capabilities of large language models.
\end{itemize}

\section{Background and Motivation}


\subsection{Inter-procedural Vulnerabilities}

Vulnerabilities often span multiple files and functions, with the vulnerability trigger location sometimes lying outside the patched function. 
According to research by Li~\etal~\cite{li2024effectiveness}, vulnerabilities frequently involve an average of 2.8 layers (or functions). 
As illustrated in Listing~\ref{lst:sg_common_write}, it is CVE-2015-8962, the vulnerability type is Double Free, and the to-be-patched function in vulnerability patch is ``sg\_common\_write''.
This issue arises because in function named ``sg\_common\_write'', a callee function named ``blk\_end\_request\_all'' releases the block request ``srp->rq'' in a control branch at line 10, yet the ``sg\_finish\_rem\_req'' function attempts to free the request object ``srp->rq->cmd'' in certain branches at line 11, which may result in freeing a memory block twice, as after ``blk\_end\_request\_all'' is called, ``srp->rq'' points to an invalid memory area. 
However, in ``sg\_finish\_rem\_req'', ``srp->rq->cmd'' is released again, and this pointer has been released when ``srp->rq'' is released.
If a vulnerability detection model only receives to-be-patched function ``sg\_common\_write'' without the concrete contents in other functions, it will be unable to find the root cause of the vulnerability and may fail to identify the vulnerability or report the wrong problem type.
We test this vulnerability on ChatGPT-4o. 
The model responds:
``While these issues do not necessarily constitute a confirmed vulnerability directly, they are indeed potential security risks. To fully assess its severity, you also need to understand:
The context in which this function is called and its inputs can be controlled by an attacker and whether the code is running in a kernel or sensitive environment with high privileges.''

However, providing the complete code for these functions will introduce excessive noise, hindering the model's ability to identify relevant information about the vulnerability. 
Therefore, a more in-depth analysis of callee functions of to-be-patched functions in vulnerability patches is essential to discover the root cause of the vulnerability.
Furthermore, after analyzing this vulnerability, we find that resource-related vulnerabilities, whether cross-function or not, are often closely linked to certain key resource management APIs. 
In this example, it is evident that the vulnerability is associated with the fundamental memory management API, ``malloc(), free()''. 
Therefore, we attempt to identify APIs involved in critical resource operations to assist in resource-related vulnerability detection, and then further expand to the detection of other types of vulnerabilities.

\begin{center}
\vspace{2mm}
\scalebox{0.75}{ 
\begin{lstlisting}[
		language=C++,
		caption={A Double Free Vulnerability Example (CVE-
2015-8962)},
		label={lst:sg_common_write}
	]
static int sg_common_write(Sg_fd * sfp, Sg_request * srp,
                unsigned char *cmnd, int timeout, int blocking){
    ...
    k = sg_start_req(srp, cmnd);
    if (k) {
    ...
    }
    if (atomic_read(&sdp->detaching)) {
        if (srp->bio)
            blk_end_request_all(srp->rq, -EIO);
        sg_finish_rem_req(srp);
        return -ENODEV;}
    ...
static int sg_finish_rem_req(Sg_request *srp) {
    ...
    if (srp->rq) {
        if (srp->rq->cmd != srp->rq->__cmd)
            free(srp->rq->cmd);
        blk_put_request(srp->rq); }

    ... }
void blk_end_request_all(struct request *rq, blk_status_t error) {
    ...
    if (unlikely(blk_bidi_rq(rq)))
            bidi_bytes = blk_rq_bytes(rq->next_rq);
    blk_finish_request(rq, error);
    ... }
void blk_finish_request(struct request *req, int error){
    ...
    __blk_put_request(req->q, req);
}
static inline void __blk_free_request(struct request_list *rl, struct request *rq)
{
    ...
    mempool_free(rq, rl->rq_pool);
}
void mempool_free(void *element, mempool_t *pool)
{
    ...
    pool->free(element, pool->pool_data);
}

\end{lstlisting}
 }
\vspace{4mm}
\end{center}

\subsection{Limitations of Existing Vulnerability Detection Methods}

With the rise of LLMs, there have been numerous attempts to leverage them for vulnerability detection.
Unfortunately, we observe that existing approaches~\cite{zhang-etal-2023-black, zhang2024prompt, steenhoek2024comprehensive} all failed to detect this vulnerability.
By further investigating the underlying reason, Zhang~\etal~\cite{zhang-etal-2023-black} found that LLMs generally learn shallow information about vulnerabilities, such as token meanings, and tend to misjudge when certain tokens are modified, such as function names. 
Besides, Zhang~\etal~\cite{zhang2024prompt} found that large language models can easily detect syntax-related or boundary-related vulnerabilities, but they fail to achieve high accuracy for vulnerabilities that require a complete understanding of the context. 
Therefore, they tested the impact of data flow and API call information as supplementary information for vulnerability code on model detection performance and found that adding certain contextual information is effective for large models.
Similarly, Steenhoek~\etal~\cite{steenhoek2024comprehensive} conducted empirical research on LLMs for vulnerability detection, they found that the performance was close to random guessing, with accuracy rates between 50\% and 60\%. 
Thus, it is clear that the effectiveness of vulnerability detection using LLMs is significantly constrained by the context of the vulnerability. 
Furthermore, these studies did not investigate the effectiveness of LLMs for complex inter-procedural vulnerabilities. 
If only to-be-patched functions and their internal information are used as samples for vulnerability detection, it becomes even more difficult to identify the root cause, and accurately detect vulnerabilities.

\subsection{Our approach}
We aim to use abstraction based on primitive API to represent each callee function within each to-be-patched function, simplifying each callee and highlighting the primitive API-related operations it performs. 
For instance, in the vulnerability shown in Listing~\ref{lst:sg_common_write}, we attempt to abstract the callee functions using control flow and data flow analysis.
We extract all control branches within each callee function to identify branches containing specific primitive APIs and use the conditions under which these APIs appear as concrete branch abstraction. 
We then determine whether a particular primitive API is invoked in all, some, or none of the branches, forming a fuzzy control branch abstraction for that API. 
Additionally, we count the occurrences of the primitive API within the callee function and analyze the data objects operated on by the API.
Finally, we use the fuzzy conditions, concrete branch conditions, occurrence counts, and key data objects operated on by the primitive API as an abstraction for each callee function, which we refer to as the Primitive API abstraction.

\section{Vulnerability Detection Framework}

\begin{figure*}[htb]
    \centering 
    \includegraphics[scale=0.45]{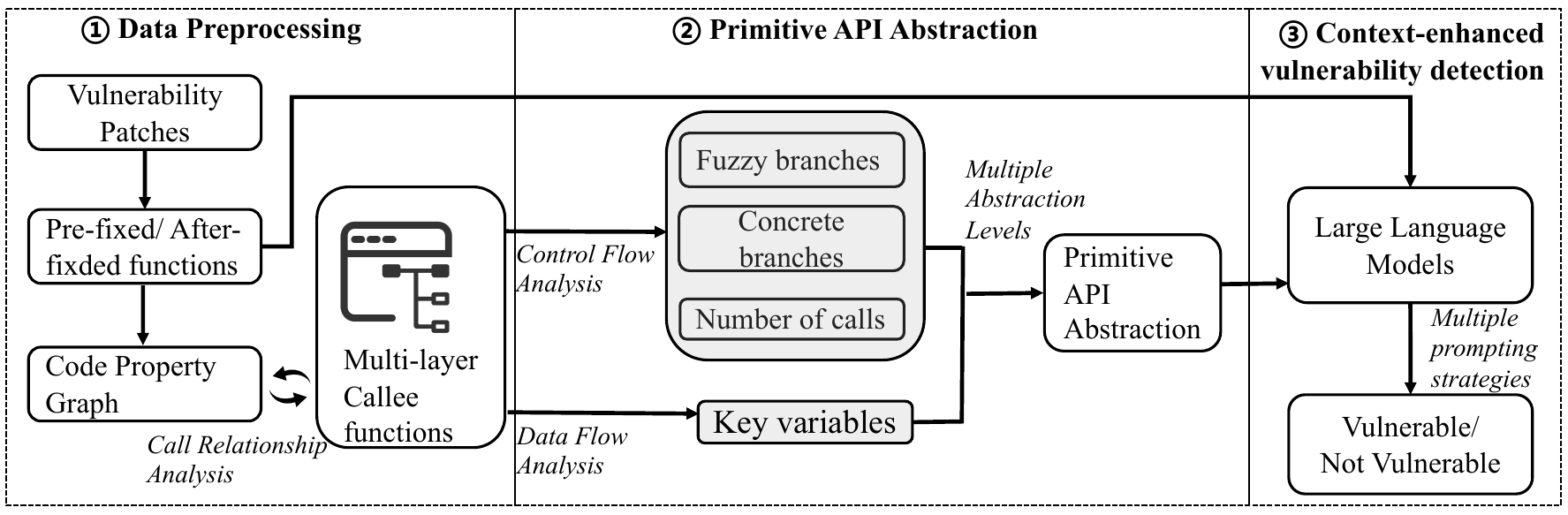}
    \vspace{2mm}
    \caption{Primitive API Abstraction and Context-Enhanced
Vulnerability Detection Framework} 
    \label{fig:framework}
\end{figure*}

We propose a context-enhanced vulnerability detection framework named \toolname that addresses the limitations of existing approaches in detecting complex inter-procedural vulnerabilities.
The input to PacVD includes a target function and its contextual information related to callee functions within a specific scope. 
In practical scenarios, the target function may be a user-specified suspicious function requiring detection, or it could originate from code changes committed to version control systems or periodic scans of entire codebases. 
The output is whether the target function contains vulnerabilities.
As illustrated in Figure~\ref{fig:framework}, our framework comprises three key components: 
(1) Data Preprocessing, responsible for extracting and preparing code for analysis; 
(2) Primitive API Abstraction, which extracts contextual information relevant to vulnerabilities; 
and (3) LLM-based Vulnerability Detection, which leverages contextual information for precise vulnerability identification.

Traditional vulnerability detection methods typically adopt a single perspective: function-level approaches lack sufficient contextual information, while repository-level methods introduce excessive noise and computational overhead. 
Our approach extracts key API behaviors through program analysis, maintaining low computational costs while preserving sufficient contextual information. 
This framework is applicable not only to historical vulnerability analysis but can also be integrated into the software development process to meet vulnerability detection needs at different stages.

\subsection{Data Preprocessing}

The first phase of the framework involves determining the code analysis scope and conducting data preprocessing.
A study by Li et al.~\cite{li2024effectiveness} indicates that inter-procedural vulnerabilities are prevalent with an average of 2.8 inter-procedural layers.
According to their statistics, nearly 75\% of the vulnerability types have an average call depth of no more than three layers
Thus, we established a default analysis depth of three function call layers, that is, we analyze the target function and all its callee functions within a three-layer call depth as our scope.
After establishing the analysis scope, the system constructs Code Property Graphs (CPGs) of relevant functions through static program analysis. 
These CPGs serve as a unified code representation, capturing multiple static properties including control flow, call graphs, and data flow. 
For each target function requiring analysis, the system constructs CPGs not only for the function itself but also callee functions. 
This multi-level analysis is crucial for capturing cross-function vulnerabilities.
The output of these steps is a collection of target functions with their calling relationships, along with corresponding code property graph representations, which provide the foundation for the next phase that follows.

\subsection{Primitive API Abstraction}
\label{sec:Primitive API Abstraction}
\subsubsection{Primitive API Selection Rationale}
To address the complexity of real-world software systems, we introduced the primitive API abstraction phase. 
In this phase, we simplify and summarize complex function calls into corresponding primitive API representations. These primitive APIs are primarily standard library functions or system calls commonly used to manage resources (such as memory, file descriptors, or network connections), as shown in Table~\ref{tab:annotated_api}.

The selection of primitive APIs is based on the work of Song~\etal~\cite{song2024provenfix}, determined through analysis of widely used C language libraries and POSIX standard APIs (such as <stdlib.h>, <stdio.h>). 
The motivation for selecting these APIs lies in their fundamental nature — primitive APIs are basic operations common across various scenarios, and vulnerabilities typically originate from the improper use of these APIs.
For example, functions like open, fopen, fdopen, and opendir are used to open files or directories, and they require corresponding closing functions like close, fclose, and closedir to ensure proper resource closure and prevent resource leaks.
Functions such as malloc, realloc, and calloc are used for memory allocation and must be paired with free to prevent memory leaks use-after-free, double free, memory leak and other memory issues.
Thus, by analyzing the presence of these Primitive APIs in function to be tested and its callee functions, certain types of vulnerabilities can be effectively detected. 
Common primitive APIs are summarized in Table~\ref{tab:annotated_api}, which highlights the different types of vulnerabilities associated with these APIs.
Inspired by the motivation of the primitive API, we expand the primitive API to a broader context, that is, all APIs related to security operations~\cite{li2022sysevr}.



\begin{table}[H]
    \centering
    \caption{Primitive API and Corresponding Vulnerability Types}
    \label{tab:annotated_api}
    \resizebox{0.67\textwidth}{!}{
    \begin{tabular}{>{\raggedright\arraybackslash}m{5.5cm}>{\centering\arraybackslash}m{4cm}}
        \toprule
        \textbf{Primitive APIs} & \textbf{Targeted Vulnerability Type} \\
        \midrule
        open/socket/fopen/fdopen/opendir/ \newline 
        close/fclose/endmntent/fflush/closedir & Resource Leak  \\
        \midrule  
        malloc/realloc/calloc/localtime & Null Pointer Dereference  \\
        \midrule  
        malloc/free & Memory Leak, Use-After-Free, Double Free  \\
        \bottomrule
    \end{tabular}
    }
\end{table}

\subsubsection{API Abstraction Methodology}


For each target function and its callee functions, we perform control flow and data flow analysis to extract API usage features from four dimensions:

\begin{itemize}[leftmargin=*]
\item \textbf{Fuzzy Branches:}
Fuzzy branches refer to representing the callee functions using the fuzzy information on whether primitive APIs are found in their control flow branches.
It provides a global overview of API usage features, reducing information noise while capturing critical control flow characteristics. 
First, we locate the target function by its unique identifier within the codebase. 
Once identified, we traverse the function's Control Flow Graph (CFG)  to find the location of its callee functions. 
This traversal is performed recursively into deeper callee functions, to ensure that all relevant nested function calls are explored, providing a comprehensive understanding of resource operations, even if the operations occur in functions several layers deep in the call hierarchy.
When we traverse to a callee function, we conduct a thorough examination to determine if it involves resource operations.
Specifically, each operation is tracked across every control flow branch within the callee function and a defined number of its nested callees, using a depth-first traversal approach.
If a resource operation is detected across all branches, we conclude that the corresponding primitive API is called in all branches, if it is detected in specific branches but not others, we conclude that the primitive API is called in only some branches, if it is not detected, the conclusion is the primitive API is not called in any branch.
Finally, the fuzzy branches summary of resource usage is then integrated into the context of the target function, allowing for a holistic understanding of resource operations, including the influence of nested function calls. 
The fuzzy branches summary is subsequently used as part of a broader analysis to detect potential vulnerabilities, such as memory leaks, double-free errors, or use-after-free issues.
For the example of Listing~\ref{lst:sg_common_write}, the fuzzy branch abstraction we extracted is as follows:

    \noindent\colorbox{lightgray}{%
        \parbox{\linewidth}{ In the ``blk\_end\_reques\_all'' function:
On all branches, the ``free'' API is called.
On no branch, the ``malloc'' API is called.
In the ``sg\_finish\_rem\_req'' function:
On some branches, the ``free'' API is called.
On no branch, the ``malloc'' API is called.
        }
    }

\item \textbf{Concrete Branches:} 
Concrete branches record the specific control conditions and precise path information of the primitive API, illustrating under what conditions the APIs are triggered. 
They are used to provide accurate conditional execution contexts.
When traversing all control flow branches of a target function, including its callee functions and nested callees up to a specific depth, if a primitive API is called in certain branches, we add the specific control conditions of those branches to the summary of the corresponding primitive API.

    \noindent\colorbox{lightgray}{%
        \parbox{\linewidth}{ In the ``blk\_end\_reques\_all'' function, the ``blk\_finish\_request'' function is called. 
        In the ``blk\_finish\_request'' function, the ``\-\-blk\_put\_request'' function is called.
        In the ``\-\-blk\_put\_request'' function, the ``mempool\_free'' function is called. In the ``mempool\_free'' function,
if unconditionally, the ``free'' API is called.
In the ``sg\_finish\_rem\_req'' function: 
If $(\text{srp} \rightarrow \text{rq})$ and $(\text{srp} \rightarrow \text{rq} \rightarrow \text{cmd} \neq \text{srp} \rightarrow \text{rq} \rightarrow \_\_cmd)$, the ``free'' API is called.
        }
    }

\item \textbf{Number of Calls:}
Number of Calls is used to quantify the frequency of each primitive API within the call chain of the target function, providing a quantitative metric for API usage. 
This metric is utilized to identify resource operation imbalances
When traversing all control flow branches of a target function, including its callee functions and nested callees up to a specific depth, we count the occurrences of each primitive API and add this information to the summary of the corresponding primitive API.

    \noindent\colorbox{lightgray}{%
        \parbox{\linewidth}{In the ``blk\_end\_reques\_all'' function:
	    the ``malloc'' API is called 0 times,
	    the ``free'' API is called 1 times.
     In the ``sg\_finish\_rem\_req'' function:
	    the ``malloc'' API is called 0 times,
	    the ``free'' API is called 1 times.
        }
    }

\item \textbf{Key Variables:}
Key Variables are used to identify specific variables or data objects affected by primitive API operations, tracking the relationships between APIs and variables. 
This approach can be leveraged to detect variable-level vulnerabilities.
When traversing all control flow branches of a specified function, including its callee functions and nested callees up to a specific depth, if a primitive API is detected, we add the identifier of the variable being operated on to the summary. 
This is achieved through a combination of control flow and data flow analysis.

    \noindent\colorbox{lightgray}{%
        \parbox{\linewidth}{In the ``blk\_end\_reques\_all'' function:
	    the ``free'' API operates on the ``$(\text{srp} \rightarrow \text{rq})$'' variable.
In the ``sg\_finish\_rem\_req'' function:
	    the ``free'' API operates on the ``srp'' variable.
        }
    }

\end{itemize}

\subsubsection{Primitive API Abstraction Levels}
We designed four different levels of Primitive API abstraction, as shown in Table~\ref{tab:AbsLevel}. 
These abstraction levels are used to analyze the impact of different Primitive API information on vulnerability detection.






\begin{table}[H]
    \centering
    \footnotesize
    \caption{Primitive API Abstraction Levels}
    \label{tab:AbsLevel}
    \resizebox{0.9\textwidth}{!}{
    \begin{tabular}{m{2.3cm} m{10cm}}
        \toprule
        \textbf{Abstraction Level} & \textbf{Description} \\
        \midrule
        API Level 1 & This is the highest level of abstraction, using only \textit{Fuzzy Branches} summary information for different Primitive APIs. \\
        \midrule
        API Level 2 & \textit{Concrete Branches} of different Primitive APIs. \\
        \midrule
        API Level 3 & \textit{Concrete Branches} of different Primitive APIs combined with \textit{Number of Calls}. \\
        \midrule
        API Level 4 & \textit{Concrete Branches} of different Primitive APIs combined with \textit{Key Variables}. \\
        \midrule
        w/o API Level & No contextual information \\
        \bottomrule
    \end{tabular}
    }
\end{table}

The multi-dimensional API abstraction method preserves critical context while avoiding excessive details, providing high-quality input for vulnerability detection by LLMs.



\subsection{Model Prediction and Optimization}
The final stage of the framework utilizes large language models to analyze code and its context for potential vulnerabilities.
The goal is to leverage the power of LLMs to perform effective vulnerability classification, enhanced by the previous phases of context extraction.

\subsubsection{Input Representation}
After performing primitive API abstraction, we incorporate the information from different API levels as supplementary code structure and semantic details, directly appending them to the corresponding target functions. These are then jointly fed as prompts into the large language model.

\subsubsection{Prompt Engineering}
Then we utilize prompt engineering to help LLMs better understand the relationship between code and contextual information of each sample, and focus on specific resource operations that are prone to vulnerabilities.
To enhance the efficacy of our evaluation experiments, we implemented several prompting strategies that have been successfully employed in large language models for various complex tasks. 
These strategies are documented in recent literature as effective means to harness the potential of language models in domain-specific applications~\cite{shanahan2023role, wei2022chain, sun2023does, brown2020language, steenhoek2024comprehensive}.

\begin{itemize}[leftmargin=*]
    \item \textbf{Role-playing Prompt}: We configured the model to assume the role of an "expert vulnerability detection system" to provide precise, direct answers, and only elaborate explanations when necessary, enhancing the relevance and utility of the outputs for vulnerability detection tasks.
    \noindent Example:\\
    \noindent\colorbox{lightgray}{%
        \parbox{\linewidth}{%
            $\mathbf{P_r}$: You are an expert vulnerability detection system. 
            Provide precise and direct answers with explanations only when necessary.
            
        }
    }

    \item \textbf{Chain-of-Thought Prompting}: This technique facilitates the model to articulate its reasoning process, aiding in transparent decision-making. This is particularly crucial for tasks that require detailed reasoning, such as identifying and explaining code vulnerabilities.
    \noindent Examples:\\
    \noindent\colorbox{lightgray}{%
        \parbox{\linewidth}{%
            $\mathbf{P^{(chain)}_{1}}$: [CODE], [API] Based on the above code and API information, please provide a detailed summary of the code's functionality, analyze the code structure, and locate all positions where pointers are constructed and dereferenced.
        }
    }\\
    \noindent\colorbox{lightgray}{%
        \parbox{\linewidth}{%
            $\mathbf{P^{(chain)}_{2}}$: Based on your previous analysis: [Code Analysis], determine whether the code contains significant vulnerabilities. Answer 'yes' or 'no' and provide reasons if vulnerabilities are identified.
        }
    }

    \item \textbf{In-context Learning}: This strategy adapts the prompts to reflect the dialogue's context or a specific scenario, thus providing the model with a richer informational background to execute the task more effectively.
    \noindent Examples:\\
    \noindent\colorbox{lightgray}{%
        \parbox{\linewidth}{%
            $\mathbf{P^{(context)}_{1}}$: As a code reviewer, evaluate this code snippet for clarity, functionality, and maintainability. Consider also the associated API information to ensure that the control flow aligns with the intended use and structure of the code. [CODE], [API].
        }
    }\\
    \noindent\colorbox{lightgray}{%
        \parbox{\linewidth}{%
            $\mathbf{P^{(context)}_{2}}$: Based on your initial observations and the API information, make a final assessment of whether the code meets the standards for clarity, functionality, and maintainability. Respond with 'yes' if improvements are needed, or 'no' if it meets the criteria.
        }
    }

    \item \textbf{Few-shot Learning Based on Random Selection}: This strategy involves providing the model with a few examples before posing the main task, to prime it on the task's requirements. We randomly selected examples from our dataset for this purpose.
    \noindent Example:\\
    \noindent\colorbox{lightgray}{%
        \parbox{\linewidth}{%
            $\mathbf{P_f}$: Code Example 1: [CODE] API Information: [API] Output: [yes/no]; 
            Code Example 2: [CODE] API Information: [API] Output: [yes/no]. 
            Refer to the examples above, then analyze the following code snippet and associated API information[CODE], [API]. 
            Provide a detailed response on the vulnerability status of the code. 
            If the code is vulnerable, start your answer with "yes" and provide a brief explanation. 
            If not, start with "no" and explain why.
        }
    }

    \item \textbf{Few-shot Learning Based on Contrastive Pairs}: We used a set of positive and negative examples (vulnerable and non-vulnerable code snippets) to further clarify the task requirements for the model. This approach helps in distinguishing subtle differences between secure and vulnerable code.
    \noindent Example:\\
    \noindent\colorbox{lightgray}{%
        \parbox{\linewidth}{%
            $\mathbf{P_{f-c}}$: Examine the 'Before Fix' and 'After Fix' code snippets to understand the vulnerability remediation. 
            Determine if the 'Before Fix' version is vulnerable, and if so, explain how the 'After Fix' version addresses the issue. Before Fix: [CODE1], After Fix: [CODE2] Refer to these examples. 
            Now, analyze the following code snippet and API Information[CODE], [API]. 
            Respond with 'yes' if it is vulnerable, otherwise answer 'no'.
        }
    }
\end{itemize}

The outcome of LLM prediction is a binary classification (\ie vulnerable/not vulnerable), which helps to prioritize areas in code that need further manual review.
Details of model configuration are shown in Section~\ref{sec:modeconfig}.
\section{Empirical Evaluation Design}
In this section, we first introduce the details of our research questions, followed by a description of the dataset used. 
Then, we explain the different strategies adopted for experimental evaluation.
Finally, we present specific implementation details of the experiments.

\subsection{Research Questions}
The experimental evaluation aims to answer the following research questions (RQs):
\begin{itemize}[leftmargin=*]
  \item \textit{RQ1: What is the effectiveness of different levels of Primitive API summaries on vulnerability detection?}
    The impact of adding different Primitive API summaries as contextual information for vulnerability detection is unknown. 
    We aim to investigate which Primitive API information is helpful for the detection capability of LLMs.
     \item \textit{RQ2: How do different API abstraction levels perform for different types of vulnerability detection?}
     Our goal is to explore how different API abstraction levels affect the results of vulnerability detection of different CWE types, so as to gain a deeper understanding of our approach.
     
  \item \textit{RQ3: How do different LLMs and prompt engineering strategies perform on vulnerability detection?}
    We study the impact of different LLMs and prompt strategies on the detection capabilities of LLMs, and study which LLMs and strategies are helpful for primitive API summaries as context information.
  \item \textit{RQ4: How does our approach compare to existing baselines?}
    We examine whether incorporating Primitive API information can improve the LLMs' performance beyond that of other methods.
\end{itemize}

\subsection{Dataset}
In existing vulnerability detection research, datasets primarily originate from the National Vulnerability Database (NVD)~\cite{NVD}. 
For our experiments, we selected a real-world C/C++ dataset named NVD-New. 
This dataset was initially constructed by Li et al.~\cite{li2022sysevr} and further expanded by Zhang et al.~\cite{Zhang2023Compare} using vulnerable programs collected from 16 open-source projects.
The original format of the dataset consists of project repositories containing CVEs and their corresponding CVE-specific information.
We utilize the detailed information of each CVE entry such as project name, commit ID, vulnerable file name, and vulnerable function name to extract vulnerable and non-vulnerable repositories and generate CPGs for them with Joern~\cite{team_joern_2024}.
Then we start from the function before and after the vulnerability fix, traverse their callee functions within a specific layer of the function, and then generate the primitive API abstraction according to the method in Section~\ref{sec:Primitive API Abstraction}.
In this step, we screen out specific memory safety vulnerabilities that are applicable to our method.
Finally, the statistical results for different types of samples with primitive API summaries are shown in Table~\ref{tab:comprehensive_vulnerabilities}.

\begin{table}[H]
    \centering
    \caption{Comprehensive Statistics of Vulnerabilities in our Dataset}
    \resizebox{\textwidth}{!}{
        \begin{tabular}{l|l|c|c|c}
            \toprule
            \textbf{Vulnerability Category} & \textbf{CWE IDs} & \textbf{Vulnerable} & \textbf{Non-Vulnerable} & \textbf{Total Occurrences} \\ \midrule
            Resource Management Errors & CWE-772, CWE-400, CWE-399 & 49 & 48 & 97 \\
            Memory Corruption & CWE-119, CWE-125, CWE-787 & 110 & 53 & 163 \\
            Memory Handling Errors & CWE-401, CWE-415, CWE-416 & 70 & 70 & 140 \\
            Pointer Related Errors & CWE-476 & 111 & 112 & 223 \\
            \midrule
            \textbf{Total} & \textbf{All CWEs} & 340 & 283 & 623 \\
            \bottomrule
        \end{tabular}
    }
    \label{tab:comprehensive_vulnerabilities}
\end{table}


\subsection{LLM Configurations}
\label{sec:modeconfig}


In our research, we chose to utilize ChatGPT-4o, DeepSeek V2.5, and CodeLLaMA-34b due to their robust capabilities and proven performance in tasks related to natural language understanding and code interpretation, which are critical for effective vulnerability detection.

\begin{itemize}[leftmargin=*]
    \item \textbf{ChatGPT-4o}: As an iteration of the Generative Pre-trained Transformer series by OpenAI, GPT-4o stands out due to its advanced understanding of context and nuanced text generation capabilities. This model is particularly effective for parsing and understanding complex language patterns, which is essential for interpreting unstructured text in vulnerability reports and code comments.

    
    \item \textbf{DeepSeek}: DeepSeek-V3 and DeepSeek-R1 are advanced AI models based on the Transformer architecture, leveraging self-attention mechanisms to capture complex dependencies in multi-modal data (e.g., text, images, and audio). DeepSeek-V3 excels in high-performance tasks through large-scale pre-training and fine-tuning, while DeepSeek-R1 is optimized for real-time, resource-constrained environments with lightweight and energy-efficient designs. Both models incorporate knowledge distillation and dynamic optimization, ensuring robustness, adaptability, and scalability across diverse applications.
    
    \item \textbf{CodeLLaMA-34b}: Running on the local infrastructure, CodeLLaMA-34b employs the VLLM library, known for its versatility in handling diverse programming tasks. By utilizing this model, we can tailor the output settings to produce highly deterministic responses that are crucial for binary decisions like vulnerability detection (yes or no). 

    \begin{itemize}
        \item \textbf{Temperature = 0.1}: This setting is crucial as it reduces randomness in the model’s output, ensuring that responses are predictable and focused on the highest probability tokens, which is vital for precise vulnerability assessment.
        \item \textbf{Top\_p = 0.95}: Nucleus sampling helps maintain response quality by focusing on the top 95\% of the probability mass, balancing diversity with relevance—a key factor in nuanced code interpretation.
        \item \textbf{Max\_tokens = 512}: Although the task ultimately involves binary classification ("yes" or "no"), the use of Chain-of-Thought Prompting (CoT) and In-Context Learning (ICL) requires the model to produce detailed reasoning in the initial response. A token limit of 512 is set to allow the model to generate a comprehensive and detailed explanation in the first round of interaction, which provides the necessary context and rationale for subsequent decision-making. This design ensures that the generated reasoning is both thorough and consistent with the requirements of binary classification, while also enhancing the interpretability and transparency of the decision process.
    \end{itemize}

\end{itemize}

For ChatGPT and DeepSeek, we directly utilize their official APIs, while for CodeLLaMA, we deploy it locally and access it through the vLLM library. 
The combination of models and configurations enables us to leverage a powerful toolkit for vulnerability detection tasks, enhancing the reliability and accuracy of our research findings. 
The unique strengths of each model contribute to a comprehensive understanding of vulnerabilities, ensuring that our evaluation is not only thorough but also nuanced and highly contextualized.

\subsection{Other Setups}
\subsubsection{Baselines}

We apply different contextual granularities of each function as our baseline methods, referring to the method named VulEval proposed by Wen~\etal~\cite{wen2024vuleval}.
For each to-be-tested function in our dataset, we utilize the raw code of all callees (\ie All Callees), raw code of callees containing Primitive APIs (\ie API-guided Sampling Callees), raw code of callees with the highest code similarity (\ie Similarity-based Sampling Callees), raw code of random sampled callee(\ie Random Sampling Callees) and raw code of hierarchically sampled callees (\ie Hierarchy Sampling Callees) as baselines, all baselines are extracted within three iteration layers.
\begin{itemize}[leftmargin=*]
    \item \textbf{All Callees:} The strategy performs a comprehensive analysis of the entire function call chain, traversing all possible paths in the call graph starting from the target function. This approach captures the complete context of function interactions by recursively analyzing each called function up to a specified depth.
    \item \textbf{API-guided Sampling Callees:} The sampling strategy leverages domain knowledge about Primitive APIs to make informed selections from the function call chain. It prioritizes functions containing specific API calls that are known to be security-sensitive or particularly relevant to the analysis context. 
    \item \textbf{Similarity-based Sampling Callees:} The strategy utilizes code similarity metrics to select three callee functions that are most relevant to the target function. It implements a multi-metric similarity analysis approach, including Jaccard similarity~\cite{rinartha2017comparative} for token-based comparison, API call pattern similarity, Edit distance~\cite{chen2004marriage} for structural similarity, BM25 text similarity~\cite{robertson2009probabilistic} for content-based comparison. 
    \item \textbf{Random Sampling Callees:} This strategy randomly selects three functions from the target function's callee functions as the context of the target function.
    \item \textbf{Hierarchy Sampling Callees:} The sampling strategy implements a level-aware selection strategy considering the structural organization of the call graph. It distributes the sampling quota across different call depths, ensuring representation from each level of the call hierarchy.
    
\end{itemize}
\subsubsection{Metrics}

We use commonly used metrics in machine learning, including Accuracy, Precision, Recall, F1-score, and Matthews Correlation Coefficient~(MCC).

\subsubsection{Implementation Details}

The experiments were conducted on four Linux servers, each equipped with two NVIDIA A100-SXM4-80GB GPUs and two Intel(R) Xeon(R) Gold 5218 CPU @ 2.30GHz CPUs with 128G memory.
For the experimental environment, we use Python 3.10.14, vllm 0.6.2 for our experiment.

\section{Evaluation Results}

\subsection{RQ1: Results of Different Primitive API Abstraction Levels}

\subsubsection{Performance Analysis Under Different API Abstraction Levels}

As shown in Table~\ref{tab:rq1_all_api}, in the absence of API information, models generally demonstrate weaker performance, with average F1 scores ranging from 42.00\% to 44.00\%. 
Interestingly, simpler prompting strategies like Basic Prompt and role-playing Prompt typically outperform more complex approaches in this environment. 
However, even in this challenging scenario, the DeepSeek-R1 model exhibited exceptional adaptability, achieving comparatively strong results despite the lack of API information.

The A1 level, providing fuzzy branch information about API calls, delivered noticeable performance improvements over the no-API baseline. 
In-context prompting emerged as the standout strategy at this level, achieving an average F1 score of 58.91\%. 
The performance gaps between different models became more pronounced, suggesting that API information amplifies the inherent capability differences between models. 
These results confirm that even basic control flow information adds significant value to vulnerability detection tasks.

At the A2 level, which introduces concrete API call branch conditions, accuracy reached its peak, indicating that specific branch information helps establish clearer decision boundaries. 
Model-strategy combination performance exhibited greater differentiation, with F1 scores ranging from as low as 6.82\% to as high as 68.63\%. Notably, the CodeLLaMA-34b model demonstrated exceptional performance at this level, achieving an F1 score of 67.48\% with the Basic Prompt strategy. These results highlight the substantial value of detailed control flow information for vulnerability detection.

The A3 level, which augments concrete branch conditions with API call frequency information, delivered the best overall performance. 
Multiple model-strategy combinations achieved their peak performance at this level, with the DeepSeek-R1+CoT combination reaching the highest F1 score of 69.09\%. 
This level appears to represent an optimal balance point between information richness and noise, effectively combining control flow semantics with lightweight statistical features.

Despite incorporating additional variable information, the A4 level showed slightly reduced performance compared to A3. 
The finding indicate that while data flow information increases complexity, its value may vary depending on model capabilities and specific task characteristics.

\begin{table}
\centering
\caption{Experimental Results (values in \%)}
\resizebox{0.9\textwidth}{!}{
\begin{tabular}{l|cccc|cccc|cccc|cccc}
\toprule
\multirow{2}{*}{Strategy} & \multicolumn{4}{c}{ChatGPT-4o} & \multicolumn{4}{c}{CodeLLaMA-34b} & \multicolumn{4}{c}{DeepSeek-V3} & \multicolumn{4}{c}{DeepSeek-R1} \\
\cmidrule(lr){2-5} \cmidrule(lr){6-9} \cmidrule(lr){10-13} \cmidrule(lr){14-17}
 & Acc & Pre & Rec & F1 & Acc & Pre & Rec & F1 & Acc & Pre & Rec & F1 & Acc & Pre & Rec & F1 \\
\midrule
\multicolumn{17}{c}{\textbf{without API Level (w/o)}} \\
\midrule
BP & 44.07 & 43.84 & 93.02 & 59.59 & 47.27 & 44.64 & 75.00 & \textbf{55.97} & 49.48 & 46.15 & 83.72 & 59.50 & 51.34 & 47.15 & 83.22 & 60.19 \\
RP & 46.39 & 44.67 & 87.79 & 59.22 & 53.25 & 45.45 & 23.26 & 30.77 & 52.16 & 47.32 & 87.39 & \textbf{61.39} & 49.44 & 47.22 & 82.93 & 60.18 \\
CoT & 47.94 & 43.64 & 59.88 & 50.49 & 54.38 & 43.24 & 9.30 & 15.31 & 45.23 & 39.52 & 46.15 & 42.58 & 44.25 & 44.15 & 98.05 & 60.89 \\
IC & 43.81 & 44.04 & 98.84 & \textbf{60.93} & 48.21 & 42.59 & 42.07 & 42.33 & 43.25 & 43.25 & 100.00 & 60.39 & 45.09 & 45.09 & 100.00 & \textbf{62.15} \\
FSR & 47.42 & 44.20 & 70.93 & 54.46 & 52.21 & 39.66 & 13.37 & 20.00 & 56.73 & 46.88 & 10.27 & 16.85 & 48.31 & 46.03 & 87.42 & 60.30 \\
FSC & 45.74 & 44.24 & 84.88 & 58.17 & 53.51 & 42.55 & 11.63 & 18.26 & 53.01 & 35.42 & 11.97 & 17.89 & 49.57 & 46.23 & 92.76 & 61.71 \\
\midrule
\multicolumn{17}{c}{\textbf{API Level 1 (A1)}} \\
\midrule
BP & 47.37 & 44.00 & 64.71 & 52.38 & 53.64 & 47.83 & 32.35 & 38.60 & 56.52 & 50.88 & 26.13 & 34.52 & 50.91 & 47.64 & 80.80 & 59.94 \\
RP & 45.07 & 43.29 & 73.53 & 54.50 & 54.97 & 50.00 & 43.38 & 46.46 & 54.76 & 47.87 & 40.91 & 44.12 & 49.82 & 47.03 & 81.75 & 59.71 \\
CoT & 44.74 & 42.52 & 66.91 & 52.00 & 54.97 & 48.65 & 13.33 & 20.93 & 47.74 & 39.85 & 47.32 & 43.27 & 44.53 & 44.09 & 100.00 & 61.20 \\
IC & 45.07 & 44.85 & 99.26 & \textbf{61.78} & 53.43 & 47.97 & 47.58 & 47.77 & 44.19 & 44.14 & 99.12 & \textbf{61.08} & 46.47 & 45.83 & 99.18 & \textbf{62.69} \\
FSR & 50.00 & 46.58 & 80.15 & 58.92 & 48.01 & 45.16 & 72.06 & \textbf{55.52} & 54.27 & 47.22 & 26.15 & 33.66 & 51.33 & 47.18 & 88.64 & 61.58 \\
FSC & 43.75 & 42.42 & 72.06 & 53.41 & 51.50 & 44.90 & 32.35 & 37.61 & 50.34 & 42.86 & 32.06 & 36.68 & 49.82 & 46.59 & 92.80 & 62.03 \\
\midrule
\multicolumn{17}{c}{\textbf{API Level 2 (A2)}} \\
\midrule
BP & 55.77 & 54.87 & 77.50 & 64.25 & 63.45 & 63.95 & 71.43 & \textbf{67.48} & 50.00 & 55.00 & 13.75 & 22.00 & 57.69 & 56.36 & 77.50 & 65.26 \\
RP & 53.85 & 53.03 & 87.50 & 66.04 & 59.31 & 63.24 & 55.84 & 59.31 & 57.69 & 65.22 & 37.50 & 47.62 & 57.05 & 55.96 & 76.25 & 64.55 \\
CoT & 59.62 & 57.80 & 78.75 & 66.67 & 50.00 & 53.33 & 20.00 & 29.09 & 60.90 & 60.44 & 68.75 & 64.33 & 52.29 & 51.70 & 97.44 & \textbf{67.56} \\
IC & 51.28 & 51.28 & 100.00 & \textbf{67.80} & 53.02 & 54.05 & 52.63 & 53.33 & 52.59 & 52.24 & 100.00 & \textbf{68.63} & 49.01 & 49.65 & 93.42 & 64.84 \\
FSR & 50.00 & 51.06 & 60.00 & 55.17 & 47.59 & 50.47 & 70.13 & 58.70 & 45.70 & 33.33 & 3.80 & 6.82 & 56.41 & 54.69 & 87.50 & 67.31 \\
FSC & 48.08 & 49.62 & 82.50 & 61.97 & 59.31 & 65.00 & 50.65 & 56.93 & 51.00 & 57.50 & 28.75 & 38.33 & 53.85 & 53.08 & 86.25 & 65.71 \\
\midrule
\multicolumn{17}{c}{\textbf{API Level 3 (A3)}} \\
\midrule
BP & 56.03 & 56.45 & 76.64 & 65.02 & 55.82 & 63.33 & 42.54 & 50.89 & 56.02 & 55.56 & 65.48 & 60.11 & 56.77 & 57.62 & 71.31 & 63.74 \\
RP & 52.53 & 53.77 & 78.10 & 63.69 & 53.01 & 61.33 & 34.33 & 44.02 & 50.58 & 56.94 & 29.93 & 39.23 & 56.92 & 56.83 & 77.61 & 65.62 \\
CoT & 55.25 & 55.73 & 78.10 & 65.05 & 46.69 & 50.00 & 13.14 & 20.81 & 54.42 & 55.83 & 59.82 & 57.76 & 53.42 & 53.77 & 96.61 & \textbf{69.09} \\
IC & 52.53 & 52.94 & 98.54 & \textbf{68.88} & 52.74 & 57.14 & 38.71 & 46.15 & 52.78 & 53.30 & 97.41 & \textbf{68.90} & 53.10 & 53.46 & 95.87 & 68.64 \\
FSR & 47.08 & 50.41 & 44.53 & 47.29 & 51.00 & 53.49 & 68.66 & \textbf{60.13} & 48.56 & 100.00 & 3.85 & 7.41 & 57.59 & 57.95 & 74.45 & 65.18 \\
FSC & 55.08 & 56.55 & 69.34 & 62.30 & 52.21 & 61.90 & 29.10 & 39.60 & 52.36 & 62.96 & 25.19 & 36.00 & 51.91 & 54.01 & 78.91 & 64.13 \\
\midrule
\multicolumn{17}{c}{\textbf{API Level 4 (A4)}} \\
\midrule
BP & 52.74 & 51.03 & 69.72 & 58.93 & 49.83 & 45.83 & 15.49 & 23.16 & 52.74 & 55.56 & 14.08 & 22.47 & 54.37 & 52.44 & 69.92 & 59.93 \\
RP & 54.11 & 51.83 & 79.58 & 62.78 & 55.87 & 57.28 & 42.45 & 48.76 & 55.20 & 60.00 & 30.22 & 40.19 & 57.64 & 54.15 & 79.86 & 64.53 \\
CoT & 54.45 & 52.41 & 69.01 & 59.57 & 49.83 & 45.83 & 15.49 & 23.16 & 57.79 & 57.89 & 54.55 & 56.17 & 51.33 & 50.20 & 99.22 & \textbf{66.67} \\
IC & 48.63 & 48.61 & 98.59 & \textbf{65.12} & 48.20 & 45.63 & 34.81 & 39.50 & 48.15 & 48.75 & 97.50 & \textbf{65.00} & 48.02 & 48.54 & 93.55 & 63.91 \\
FSR & 47.60 & 46.50 & 51.41 & 48.83 & 49.47 & 49.26 & 71.94 & \textbf{58.48} & 51.79 & 63.64 & 5.07 & 9.40 & 51.37 & 50.00 & 78.87 & 61.20 \\
FSC & 45.89 & 46.12 & 66.90 & 54.60 & 51.60 & 52.17 & 25.90 & 34.62 & 53.38 & 55.00 & 24.09 & 33.50 & 55.08 & 53.05 & 88.28 & 66.28 \\
\bottomrule
\end{tabular}
}
\parbox{\textwidth}{\footnotesize \textit{Note}: BP denotes basic prompt, RP denotes role-based prompt, CT denotes Chain-of-thought, IC denotes In-context Learning, FR denotes few-shot learning with random selected examples, FC denotes few-shot learning with contrastive pair examples.
Acc, Pre, Rec, F1 denotes Accuracy, Precision, Recall, F1 score.
}
\label{tab:rq1_all_api}
\end{table}


\subsubsection{API Level Impact on Vulnerability Detection Performance}

\begin{figure*}[htb]
    \centering 
    \includegraphics[scale=0.27]{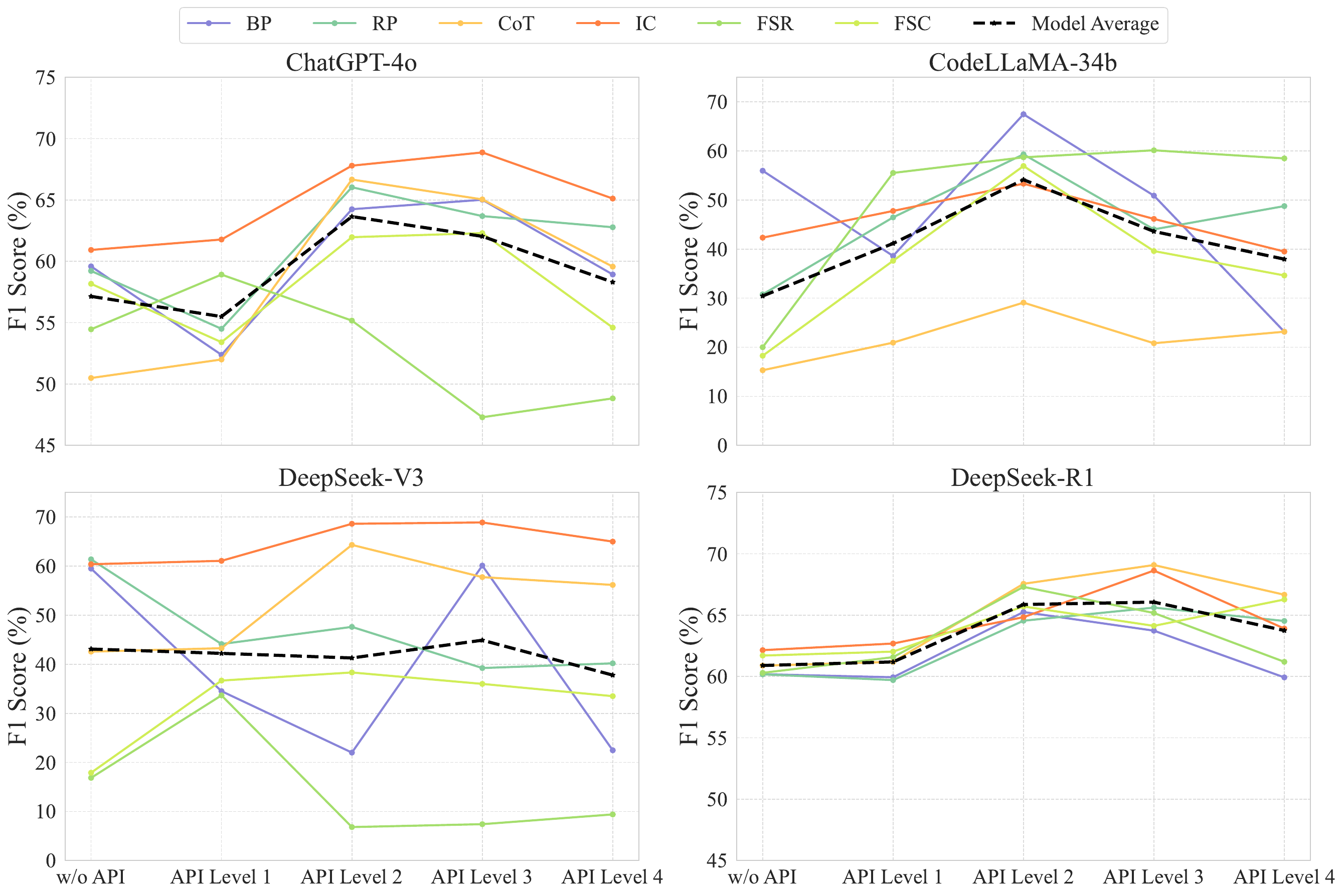}
    \vspace{2mm}
    \caption{F1 score for vulnerability detection using different API abstraction levels as auxiliary information. 
    The horizontal axis represents different levels of API abstraction.
    The vertical axis indicates the F1-score of the model for vulnerability detection. The curves in different colors represent distinct prompt engineering strategies.
    BP denotes basic prompt, RP denotes role-based prompt, CT denotes Chain-of-thought, IC denotes In-context Learning, FR denotes few-shot learning with random selected examples, FC denotes few-shot learning with contrastive pair examples.
    Model Average represents the average across all prompt strategies of a specific model.
    } 
    \label{fig:rq1_api}
\end{figure*}

As shown in Figure~\ref{fig:rq1_api}, the experimental results reveal a distinct pattern in the relationship between API abstraction levels and vulnerability detection performance. 
From no API information to A3 level, we observed a consistent upward trend in model performance, followed by a slight decline at A4 level. 
Even the most basic API abstraction (\ie A1) provided substantial performance improvements over baseline (\ie w/o API Level), with average F1 scores increasing by approximately 5\%-8\%. 
This underscores the critical importance of API contextual information for effective vulnerability detection. 
The A3 level, which combines concrete branch conditions with call frequency statistics, emerged as the optimal abstraction point, achieving the highest average F1 score across various model-strategy combinations. 
Interestingly, the A4 level, despite incorporating additional variable information, showed diminishing returns, suggesting the existence of an information-noise threshold beyond which additional details become counterproductive.

The various evaluation metrics demonstrated different sensitivities to API abstraction levels. 
Accuracy measurements peaked at the A2 level, suggesting that concrete branch information alone establishes clearer decision boundaries. 
We also observed a consistent precision-recall trade-off at higher API levels (\ie A3, A4), where models achieved improved precision at the cost of reduced recall, indicating a shift toward more conservative vulnerability identification as API information becomes more detailed.

\subsubsection{API Level Selection}

Experimental results consistently demonstrate that the A3 level (\ie specific branches + call frequency) provides the optimal cost-performance ratio and performance equilibrium point in vulnerability detection. 
The advantages exhibited by A3 can be attributed to three fundamental mechanisms: it likely represents an information saturation point where models acquire adequate structured information while additional data yields diminishing marginal returns; it establishes a computational efficiency equilibrium between model processing capability and information complexity, enabling efficient processing of all provided information; and it optimizes representation space compatibility, aligning well with most models' internal representation structures, thus facilitating effective integration of information into the reasoning process.

Besides, the relationship between path sensitivity and abstraction precision provides a theoretical foundation for understanding API abstraction effectiveness. 
A1 (\ie fuzzy branches) essentially performs path-insensitive API call abstraction, while A2-A4 maintain varying degrees of path sensitivity, with experimental results confirming that path-sensitive abstractions (\ie A2/A3/A4) significantly outperform path-insensitive ones (\ie A1). 
According to Abstract Interpretation theory, abstraction precision correlates positively with path sensitivity, though a precision-efficiency trade-off exists, with A3 achieving the optimal balance between path sensitivity and abstraction level.
This methodologically indicates that API abstraction mechanisms should prioritize control flow path differentiation capabilities rather than adding more dimensional information, supporting the applicability of "abstraction precision preservation theory" in vulnerability detection. 
Regarding context sensitivity, A2 (\ie specific branches) provides context-insensitive API call conditions, A3 adds frequency information, while A4 introduces variable associations, increasing context sensitivity to some degree. 
The superior performance of A3 over A4 suggests that lightweight contextual information is more effective than fully context-sensitive analysis.
Futhermore, the performance decline at A4 (\ie including variable information) likely stems from the inherent limitations of static alias analysis, with static analysis struggling to precisely handle complex pointer alias relationships, introducing erroneous associations in variable-level API abstractions. 

\intuition{\textbf{RQ1:} 
API integration demonstrates universal benefits, with measurable performance improvements observed even at the lowest abstraction level (A1).
A3 level achieves optimal cost-effectiveness, outperforming other tiers in comprehensive metrics.
A2 level remains highly competitive, occasionally surpassing A3 in specific model-strategy combinations.
However, adding more details (A4) leads to a slight drop in performance, indicating that there is a tipping point between information and noise.
}


\subsection{RQ2:The Impact of API Abstraction Information on Different Types of Vulnerabilities}

\begin{figure}[htbp]
    \centering
    
    \begin{subfigure}[t]{0.4\textwidth}  
        \includegraphics[width=\textwidth, height=1.5cm, keepaspectratio]{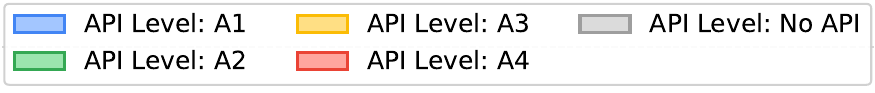}
        \label{fig:legend}
    \end{subfigure}
    
    \vspace{-0.3em}  
    \begin{subfigure}[t]{0.49\textwidth}
        \includegraphics[width=\textwidth, height=5cm]{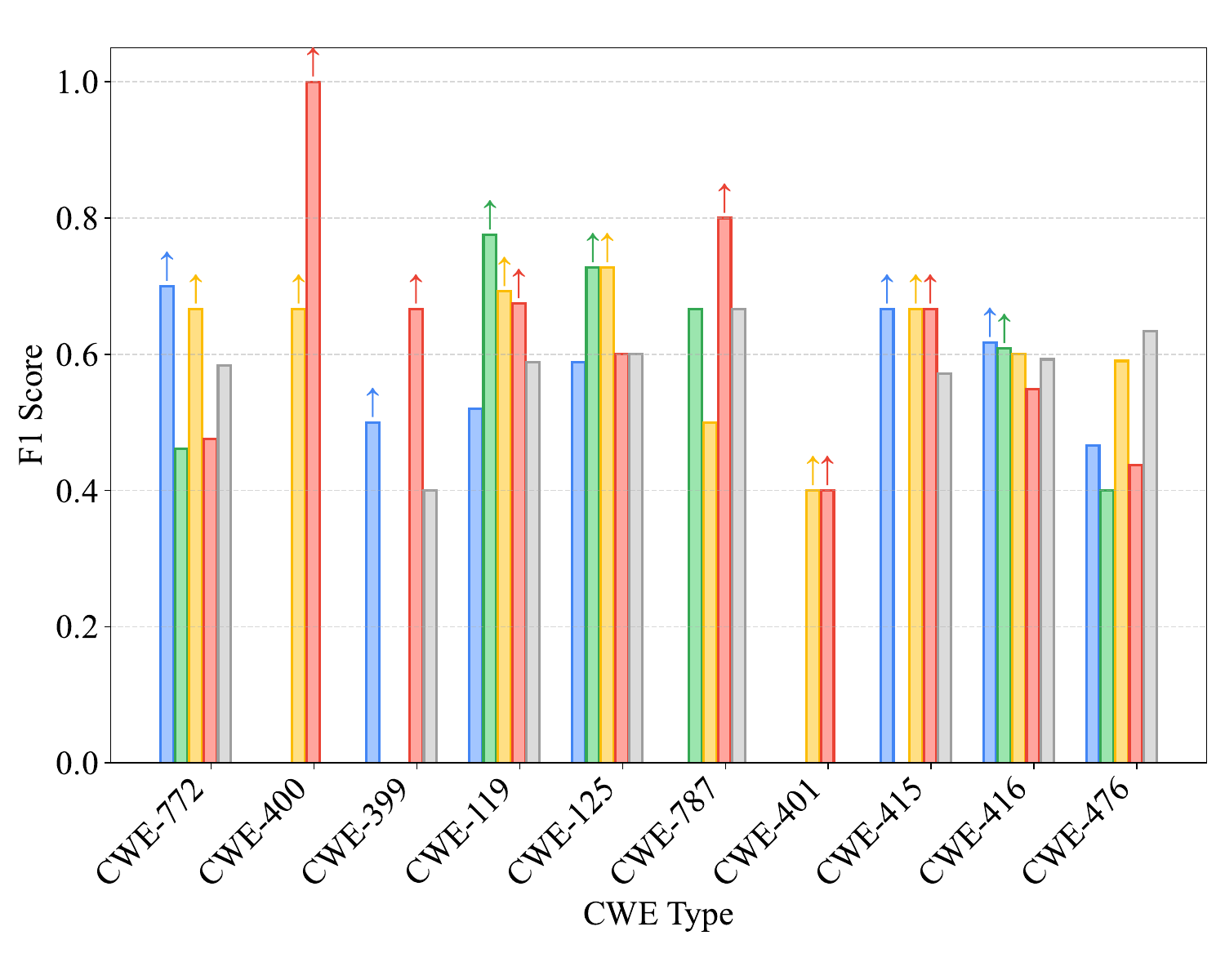}
        \caption{ChatGPT-4o}
        \label{fig:sub1}
    \end{subfigure}
    \hfill
    \begin{subfigure}[t]{0.49\textwidth}
        \includegraphics[width=\textwidth, height=5cm]{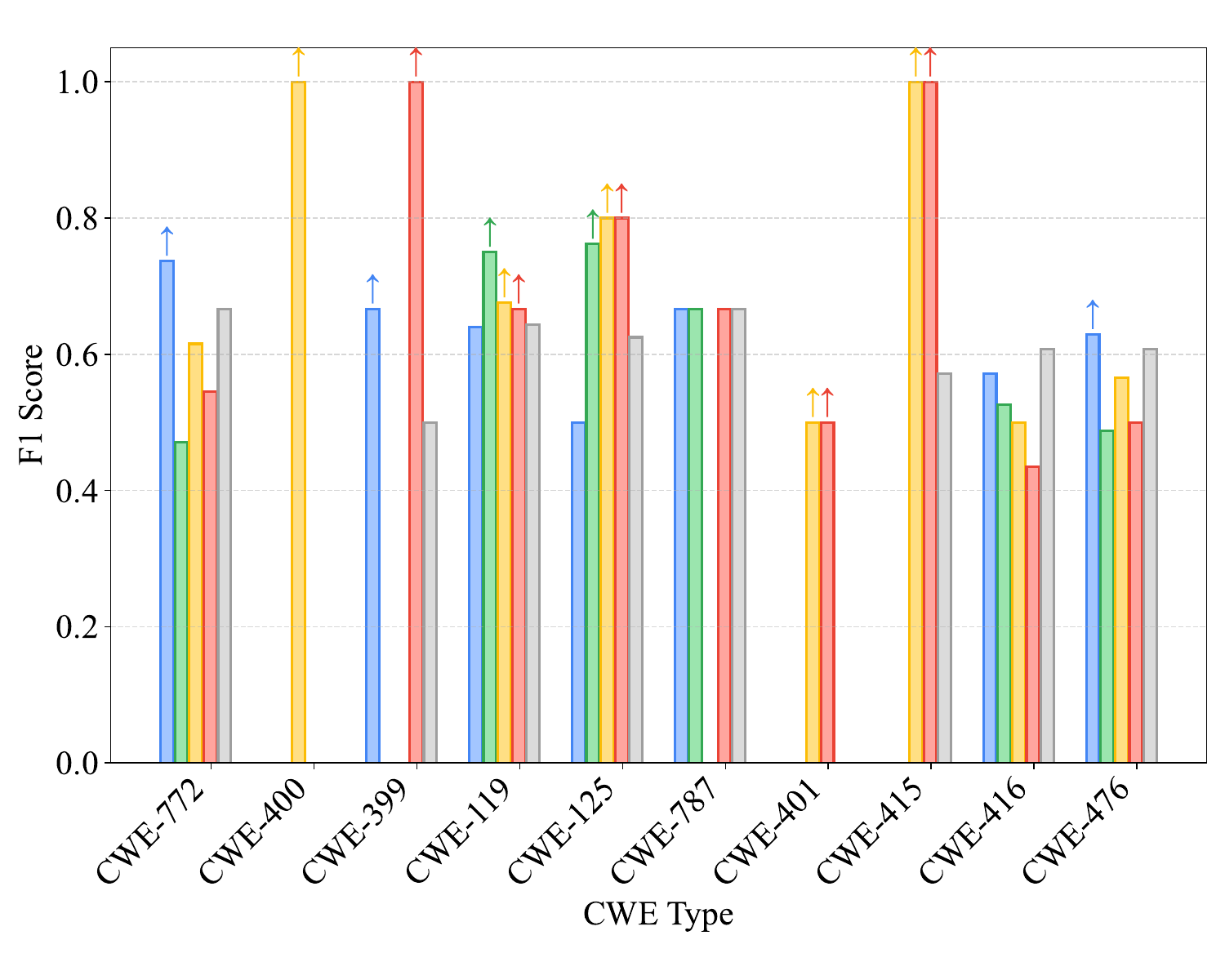}
        \vspace{-8pt}
        \caption{Deepseek-R1}
        \label{fig:sub2}
    \end{subfigure}
    
    \vspace{-0.3em}  
    \begin{subfigure}[t]{0.49\textwidth}
        \includegraphics[width=\textwidth, height=5cm]{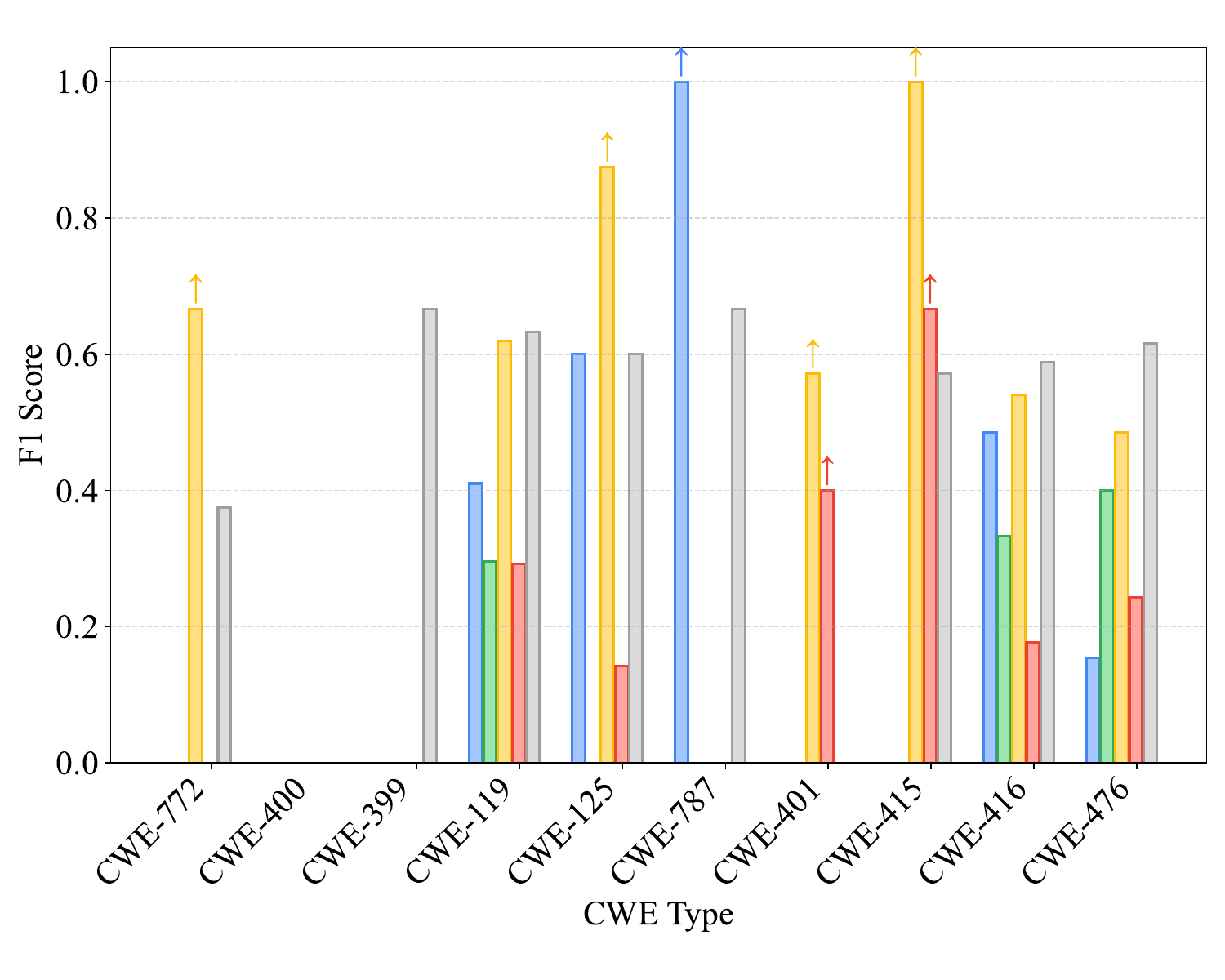}
        \vspace{-8pt}
        \caption{DeepSeek-V3}
        \label{fig:sub3}
    \end{subfigure}
    \hfill
    \begin{subfigure}[t]{0.49\textwidth}
        \includegraphics[width=\textwidth, height=5cm]{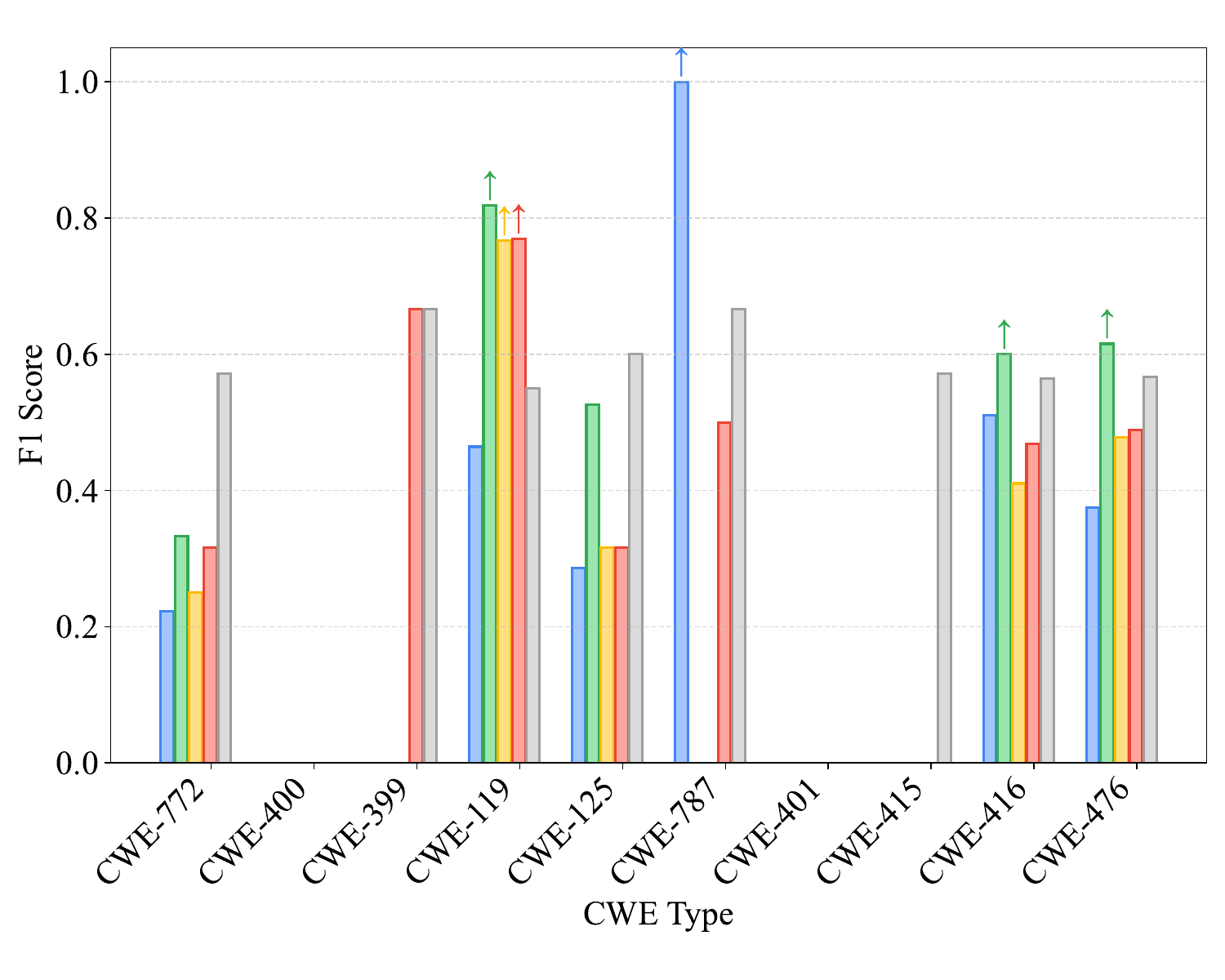}
        \vspace{-8pt}
        \caption{CodeLlama-34b}
        \label{fig:sub4}
    \end{subfigure}

    \vspace{1ex}  
    \caption{The effect of different API abstraction levels on different vulnerability types, Bar height represents F1 score. Higher is better. Arrows ($\uparrow$) indicate significant improvements over No API baseline.} 
    \label{fig:model_comparison}
\end{figure}

As shown in Figure~\ref{fig:model_comparison}, the experimental results demonstrate substantial improvements in boundary-related memory vulnerability detection when applying our API abstraction methodology. 
For CWE-787 (\ie Out-of-bounds Write), API Level A1 achieved remarkable performance in both DeepSeek-V3 and CodeLlama-34b models, with F1 scores approaching 100.00\%, representing approximately a 67.00\% improvement over the No API baseline. 
This significant enhancement indicates that even minimal API abstraction can substantially improve the detection of critical out-of-bounds write vulnerabilities.
For CWE-125 (\ie Out-of-bounds Read), multiple abstraction levels demonstrated effectiveness across various models. 
In DeepSeek-R1, DeepSeek-V3, and ChatGPT-4o, both API Level A2 and A3 consistently outperformed the No API baseline. 
Notably, in DeepSeek-V3, API Level A3 achieved an F1 score of approximately 88.20\%, representing a 47.86\% improvement over the No API baseline. 
This finding suggests that moderate levels of API abstraction can effectively capture the contextual patterns necessary for identifying out-of-bounds read vulnerabilities.
Regarding CWE-119 (\ie Improper Restriction of Operations within the Bounds of a Memory Buffer), our analysis revealed that the CodeLlama-34b model particularly benefited from API abstraction.
API Level A2, A3, and A4 all substantially outperformed the No API baseline, with improvements ranging from 43.01\% to 52.20\%.

\intuition{\textbf{RQ2-1:}
The consistent effectiveness across all abstraction levels for boundary-related memory errors indicates that our approach exhibits universality in detecting memory-related vulnerabilities, regardless of the specific abstraction granularity employed.
}

For CWE-415 (\ie Double Free), the DeepSeek-R1 model demonstrated exceptional performance with API Level A3 and A4, both achieving F1 scores of approximately 97.00\%, representing a 70\% improvement over the No API baseline. 
Similarly, in DeepSeek-V3, API Level A3 demonstrated substantial improvement over the baseline.
CWE-400 (\ie Uncontrolled Resource Consumption) detection was significantly enhanced in both DeepSeek-R1 and ChatGPT-4o models when using API Level A3 and A4. 
Most notably, ChatGPT-4o with API Level A4 achieved a F1 score of approximately 100.00 for this vulnerability type. 
Additionally, for CWE-399 (\ie Resource Management Errors), DeepSeek-R1 with API Level A4 also attained a perfect F1 score of approximately 100.00\%.

\intuition{\textbf{RQ2-2:}
Resource management-related vulnerabilities particularly benefit from higher abstraction levels (A3 and A4) that provide more detailed contextual information. The enhanced performance suggests that accurate detection of these vulnerability types requires more comprehensive API call representations that explicitly model resource allocation and deallocation patterns.
}

Our experimental results reveal distinct effectiveness patterns across different API abstraction levels:
API Level A1, despite being the lowest abstraction level, demonstrated exceptional performance for specific vulnerabilities. It significantly improved detection of CWE-787 in DeepSeek-V3 and CodeLlama-34b, achieving near-perfect F1 scores. In ChatGPT-4o, A1 enhanced detection for CWE-772, CWE-415, and CWE-416, while in DeepSeek-R1, it improved results for CWE-399, CWE-772, and CWE-476. These results indicate that even minimal API abstraction can substantially improve certain vulnerability detections, particularly for well-defined patterns like out-of-bounds writes.

API Level A2 exhibited particular effectiveness for CWE-119, CWE-125, and CWE-416 vulnerabilities. In ChatGPT-4o, it improved detection for all three vulnerability types, while in DeepSeek-R1, it enhanced detection for CWE-119 and CWE-125. CodeLlama-34b benefited from A2 abstraction for CWE-119, CWE-416, and CWE-476. This pattern indicates that A2's intermediate abstraction level is particularly suitable for detecting access violations and pointer mismanagement issues.

API Level A3 demonstrated the most consistent performance improvements across multiple models and vulnerability types. It showed significant enhancements for CWE-772 in ChatGPT-4o and DeepSeek-V3, CWE-119 in ChatGPT-4o, DeepSeek-R1, and CodeLlama-34b, CWE-125 in ChatGPT-4o, DeepSeek-R1, and DeepSeek-V3, and CWE-415 across all evaluated models. This consistency suggests that A3 represents an optimal balance between abstraction and specificity for general memory safety vulnerability detection.

API Level A4, the highest abstraction level, showed targeted effectiveness for specific model-vulnerability combinations. It significantly improved detection of CWE-399, CWE-119, CWE-415, CWE-400, and CWE-787 in ChatGPT-4o, as well as CWE-399, CWE-119, CWE-125, and CWE-415 in DeepSeek-R1. Additionally, it enhanced detection of CWE-415 in DeepSeek-V3 and CWE-119 in CodeLlama-34b. This pattern suggests that highly detailed API abstractions are particularly beneficial for complex vulnerability types that require comprehensive contextual understanding.

\intuition{\textbf{RQ2-3:}
API Level A3 represents the most balanced and consistently effective abstraction level across multiple vulnerability types and models, suggesting it captures an optimal degree of contextual information without introducing excessive noise.
}

\subsection{RQ3: Effectiveness of Different LLMs and Prompt Engineering Strategies}

\subsubsection{Large Language Model Performance Analysis}

\begin{figure*}[htb]
    \centering 
    \includegraphics[scale=0.27]{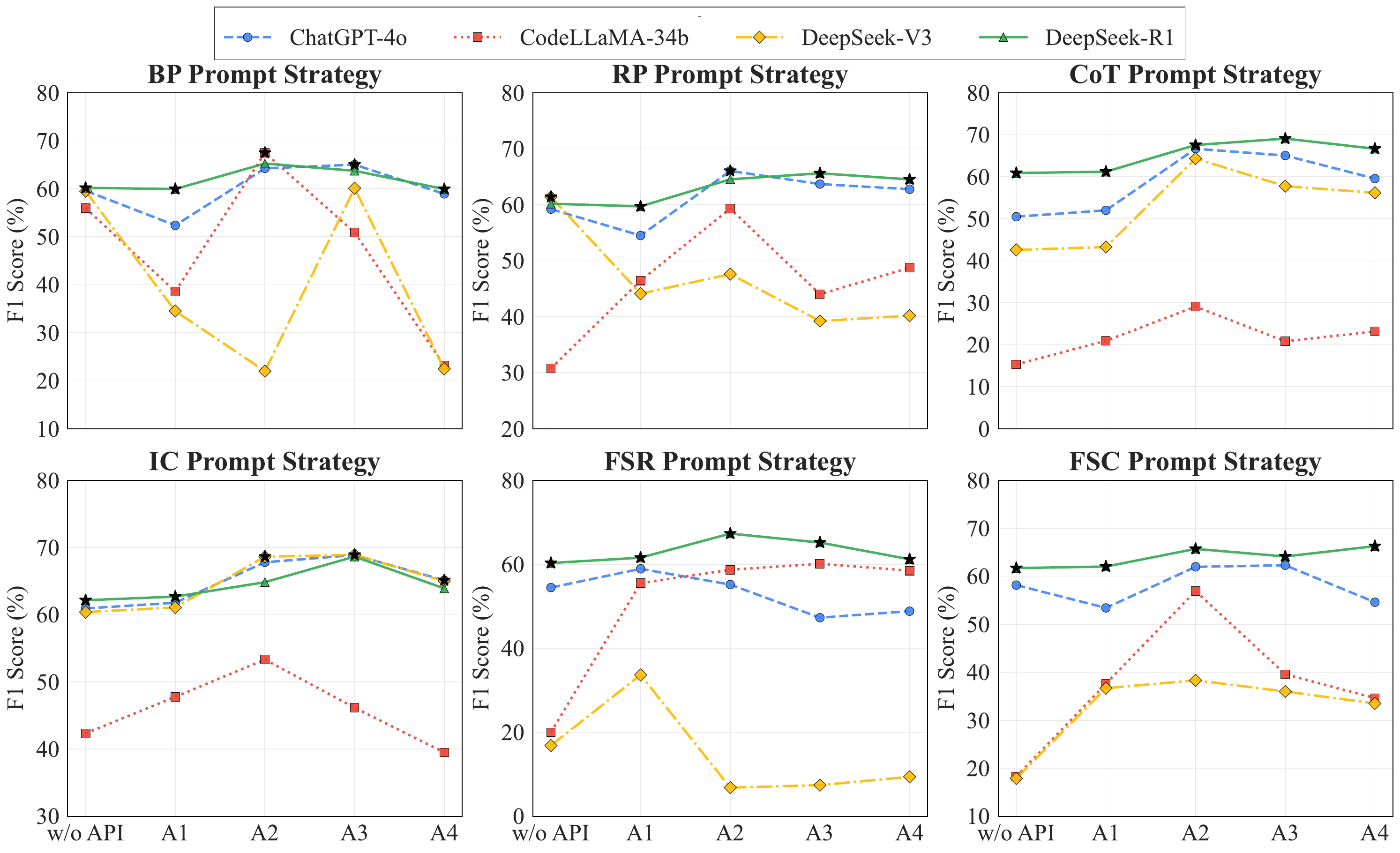}
    \caption{F1-score comparison across different vulnerability detection models.Star symbols ($\bigstar$) denote the top-performing model for each API abstraction level.} 
    \label{fig:rq3_model_compare}
\end{figure*}

As shown in Figure~\ref{fig:rq3_model_compare},
the experimental results demonstrate significant performance variations across different language models in API-assisted vulnerability detection tasks. 

DeepSeek-R1 almost outperforms all other models across all API levels. It achieves optimal performance at API level A3,  the F1 score is 69.09\%, with IC strategy proving most effective at lower API levels while CoT excels at higher levels (A2-A4).
This model maintains robust performance stability across all API information levels and prompting strategies, while maintaining exceptional recall rates throughout the evaluation. 
This superior performance can be theoretically attributed to its enhanced contextual comprehension capabilities through efficient attention mechanisms, enabling precise capture of critical information within extended sequences. 
The model's exposure to diverse pre-training data likely provided extensive experience with API-related contexts, while its balanced inference capabilities maintained high precision without sacrificing recall, indicating well-equilibrated internal representation spaces capable of flexibly adjusting decision boundaries across diverse tasks.

ChatGPT-4o demonstrates a progressive improvement in F1 scores as API levels increase, reaching its maximum performance at Level 3 , which is 68.88\%, before experiencing a slight decline at Level 4, with In-Context Learning (IC) strategy consistently outperforming all alternatives across all levels. 
ChatGPT-4o demonstrates consistently reliable performance, it maintains above-average performance at all API levels, and is second only to DeepSeek-R1 in non-API and A1 environments, particularly excelling when paired with In-context strategies and demonstrating well-balanced performance metrics.
This consistency can be attributed to diversified training objectives, with OpenAI potentially employing varied loss functions and evaluation metrics to prevent over-optimization on specific metrics at the expense of others. 
More extensive pre-training likely established a robust knowledge foundation, reducing dependence on specific API or prompting methodologies, while effective fine-tuning strategies emphasized stable performance across diverse environments.

DeepSeek-V3 achieves peak effectiveness at API Level 3, F1 score reaches 68.90\%, with IC emerging as the most efficacious strategy upon API information integration, demonstrating substantial enhancement from no-API to Level 3, followed by a modest performance reduction at Level 4. DeepSeek-V3 exhibites extreme performance volatility, displaying high sensitivity to both API levels and prompting strategies. 
While achieving exceptional results under specific configurations (\ie F1 score reaches 68.63\% with In-context strategy at A2 level), it performed poorly under others (\ie F1 score reaches 7.41\% with few-shot-random at A3 level), indicating substantial potential requiring precise calibration. 
These extreme fluctuations potentially reflect over-specialization issues, with the model potentially over-optimized for specific domains or tasks, limiting generalization capabilities. 
Its high parameter sensitivity with architectural features particularly responsive to input variations causes significant performance fluctuations from minor prompting adjustments, while unbalanced representation spaces exhibit bias toward certain prompting patterns while neglecting others.

CodeLLaMA-34b exhibits optimal performance at API Level 2, F1 score reaches 67.48\%, showing marked superiority compared to other levels, though its optimal strategy varies—Basic Prompting (BP) proves most effective at A2, while Few-Shot with Random examples (FSR) dominates at Levels 3 and 4; performance noticeably deteriorates following A2, exhibiting limitstion in recall. 
This pronounced sensitivity to API levels likely stems from its specialized training corpus; as a code-focused model, it may lack sufficient contextual comprehension in low-API environments while effectively utilizing structured information as API levels increase. Its attention mechanisms may be optimized for structured and semi-structured data analysis, while its preference for precision over recall suggests conservative internal decision thresholds.

\intuition{\textbf{RQ3-1:}
Reasoning-focused models like DeepSeek-R1 exhibit minimal sensitivity to API abstraction gradations, maintaining consistent performance across all levels and thus proving suitable for general applications.
ChatGPT-4o, with its extensive training corpus, displays moderate sensitivity with stable cross-tier performance. 
Both these two models can effectively operate with low cost and level API abstraction configurations.
Conversely, code-specialized CodeLLaMA-34b shows pronounced sensitivity, with significant improvement when transitioning from no abstraction to level A2. Similarly, DeepSeek-V3 demonstrates high sensitivity with irregular performance patterns across different tiers. 
These latter two models require more detailed API abstraction levels and more overhead to optimize vulnerability detection.
}

\subsubsection{Prompting Strategy Performance Analysis}

As shown in Figure~\ref{fig:rq1_api}, In-context Learning demonstrated superior performance across API levels with an average F1 score of 59.00\%, particularly excelling at A2 and A3 levels while maintaining high recall values. 
This effectiveness stems from direct example provision reducing model burden, implicit runtime adjustment effects, and expanded decision boundaries.

Chain of Thought (CoT) performed well only at higher API levels (A2-A4) with more capable models, reaching optimal F1 scores of 69.09\% with DeepSeek-R1 at A3. 
Performance varies significantly based on model computational capacity, with less capable combinations suffering from error accumulation in multi-step reasoning.

Role-playing maintained consistent performance across all configurations with minimal performance drops, proving most reliable especially at A4 level with DeepSeek-R1. 
This stability results from established response frameworks constraining output space and effective activation of role-relevant knowledge clusters.

Basic Prompt achieved optimal performance in no-API environments (\ie F1 score is 58.86\% and with CodeLLaMA-34b at A2 level (\ie F1 score is 67.48\%), though effectiveness decreased at higher API levels. 
In information-scarce scenarios, direct queries allow focused computational resource allocation on core reasoning rather than processing complex prompts.

Few-shot strategies performed well primarily with DeepSeek-R1, reaching 67.31\% F1 with random sampling at A2 level, but showed inconsistent results with other models. 
This variability reflects differing minimum effective sample quantity requirements and representation space compatibility issues across models.

\intuition{\textbf{RQ3-2:}
Prompt strategy effectiveness exhibits clear dependency on API abstraction level. Without API abstraction, Basic Prompt outperforms other strategies. 
At low abstraction (A1), In-context Learning demonstrates superior performance, followed by few-shot approaches. 
Medium abstraction (A2) benefits most from In-context Learning and Chain-of-Thought techniques. 
At high abstraction levels (A3-A4), both In-context Learning and Chain-of-Thought prompting yield comparable optimal results. 
These findings indicate that as contextual information richness increases through API abstraction, more sophisticated reasoning-oriented prompt strategies become advantageous.

}

\subsection{RQ4: Baseline Comparison}

\begin{table}[h]
    \centering
    \caption{Comparison of \toolname and Baselines for Vulnerability Detection.}
    \vspace{2mm}
    \footnotesize
    \renewcommand{\arraystretch}{1.2} 
    \resizebox{0.85\textwidth}{!}{
        \begin{tabular}{lccccccc}
            \toprule
            \textbf{Method} & \textbf{Metric} & \textbf{\toolname} & \textbf{AL-C} & \textbf{AS-C} & \textbf{SS-C} & \textbf{RS-C} & \textbf{HS-C} \\
            \midrule
            \multirow{5}{*}{DeepSeek-R1 + CoT} 
            & Accuracy   & \textbf{53.42}         & 50.54 & 50.00 & 47.37 & 53.26 & 49.46  \\
            & Precision  & \textbf{53.77}         & 50.00 & 49.30 & 48.86 & 50.00 & 48.81  \\
            & Recall     & 96.61         & 97.83 & 94.59 & 89.58 & 97.67 & 91.11  \\
            & F1 Score  & \textbf{69.09} & 66.18 & 64.81 & 63.24 & 66.14 & 63.57  \\
            \midrule
            \multirow{5}{*}{DeepSeek-R1 + In-Context} 
            & Accuracy  & \textbf{53.10}         & 50.00 & 52.05 & 50.00 & 51.65 & 48.94  \\
            & Precision & \textbf{53.46}         & 50.00 & 51.47 & 49.44 & 50.56 & 48.86  \\
            & Recall    & 95.87         & 93.48 & 94.59 & 100.00& 100.00& 93.48  \\
            & F1 Score  & \textbf{68.64} & 65.15 & 66.67 & 66.17 & 67.16 & 64.18 \\
            \midrule
            \multirow{5}{*}{DeepSeek-V3 + In-Context} 
            & Accuracy  & \textbf{52.78}         & 46.94 & 50.00 & 50.00 & 48.98 & 51.02 \\
            & Precision & \textbf{53.30}         & 47.62 & 50.00 & 49.41 & 48.91 & 50.00 \\
            & Recall    & \textbf{97.41}    & 83.33 & 92.50 & 87.50 & 93.75 & 93.75 \\
            & F1 Score  & \textbf{68.90}& 60.61 & 64.91 & 63.16 & 64.29 & 65.22 \\
            \midrule
            \multirow{5}{*}{GPT-4o + In-Context} 
            & Accuracy  & \textbf{52.53}         & 50.00 & 51.25 & 52.58 & 48.98 & 48.45 \\
            & Precision & \textbf{52.94}         & 48.94 & 50.63 & 51.06 & 48.98 & 48.94 \\
            & Recall    & 98.54         & 100.00 & 100.00 & 100.00 & 100.00 & 95.83 \\
            & F1 Score  & \textbf{68.88}& 65.71 & 67.23 & 67.61 & 65.75 & 64.79\\
            \midrule
            \multirow{5}{*}{CodeLLaMA-32b + Basic} 
            & Accuracy  & \textbf{63.45}         & 60.00 & 61.46 & 56.25 & 56.25 & 62.50 \\
            & Precision & 63.95       & 66.67 & 64.10 & 57.50 & 58.82 & 65.79 \\
            & Recall    & \textbf{71.43}         & 40.00 & 52.08 & 47.92 & 41.67 & 52.08 \\
            & F1 Score  & \textbf{67.48} & 50.00 & 57.47 & 52.27 & 48.78 & 58.14 \\
            \bottomrule
        \end{tabular}
    }
    \parbox{\textwidth}{\footnotesize \textit{Note}: AL-C, AS-C, SS-C, RS-C and HS-C denotes baselines on all callees setting, on API-guided sampling callees setting, on similarity-based sampling callees setting, on random sampling callees, as well as on hierarchy sampling callees setting.}
    \label{tab:comparison_methods}
\end{table}

 Our comprehensive evaluation demonstrates that PacVD consistently outperforms existing baselines across all tested model configurations for vulnerability detection, as shown in Table~\ref{tab:comparison_methods}.

PacVD achieves superior F1 scores across all model configurations, indicating a robust and generalizable approach to vulnerability detection. 
This performance advantage stems from PacVD's ability to maintain an optimal balance between precision and recall—a critical factor in practical vulnerability detection systems. 
While most baseline methods exhibit a tendency toward high recall at the expense of precision, particularly All Callees and Random Sampling approaches, PacVD consistently maintains high recall rates while simultaneously achieving improved precision.

Besides, the performance characteristics vary notably across different model configurations. 
With DeepSeek-R1 + CoT, PacVD achieves its highest F1 score, outperforming the closest baseline by 4.40\%. 
Similar performance advantages are observed with DeepSeek-R1 + In-Context and DeepSeek-V3 + In-Context configurations, the F1 score of PacVD outperform the closest baseline by 2.20\% and 5.64\%. 
Under the GPT-4o + In-Context setting, the F1 score of PacVD outperform the closest baseline by 1.88\%. 
Although multiple baselines achieve perfect recall, their precision suffers substantially, whereas PacVD provides a more balanced performance profile. 
Under the CodeLLaMA-32b + Basic configuration, the F1 score of PacVD outperform the closest baseline by 16.06\%.
Interestingly, the CodeLLaMA-32b + Basic configuration exhibits distinct characteristics, with generally higher precision but lower recall across all methods, yet PacVD still maintains the highest overall F1 score.

Our analysis of individual sampling strategies reveals important insights. 
The All Callees (AL-C) approach, while comprehensive, suffers from precision limitations despite its high recall in most configurations. 
API-guided Sampling (AS-C) demonstrates moderate stability across different model settings, while Similarity-based Sampling (SS-C) shows considerable performance variation across models, performing particularly well with GPT-4o. 
Random Sampling (RS-C), despite its simplicity, achieves surprisingly competitive results in certain configurations, often with perfect recall but compromised precision. Hierarchy Sampling (HS-C) exhibits model-dependent performance, excelling primarily with CodeLLaMA-32b but underperforming with other models.


From a practical perspective, our results indicate that PacVD offers substantial advantages for real-world vulnerability detection systems by reducing false positives while maintaining high detection rates. The consistent performance across diverse model configurations suggests that PacVD provides a robust foundation for deployment in varied software engineering environments.

\intuition{\textbf{Summary for RQ4:}
Our approach demonstrates competitive performance against existing baselines, achieving the highest Accuracy, Precision and F1 scores in all configurations while maintaining exceptional recall rates, indicating a better balance between precision and false alarm reduction that makes it particularly suitable for practical vulnerability detection scenarios.
}

\section{Discussion}

\subsection{Practical Implications for Software Security}

Our comprehensive evaluation of API abstraction levels for vulnerability detection through large language models yields several important implications and lessons for both practitioners and researchers:

\paragraph{Optimal API Abstraction Selection:} Our findings suggest that A3 level abstraction (\ie specific branches + call frequency) provides the best balance between information richness and noise for vulnerability detection. Organizations implementing LLM-based vulnerability detection should prioritize this abstraction level to maximize detection accuracy while minimizing computational overhead.

\paragraph{Model-Strategy Pairing:} Different models exhibit varying levels of sensitivity to API abstraction. For resource-constrained environments, DeepSeek-R1 offers exceptional performance with minimal sensitivity to abstraction levels, making it ideal for general application. In contrast, specialized code models like CodeLLaMA-34b require more detailed API abstraction to reach optimal performance.

\paragraph{Prompt Engineering Guidelines:} As API contextual information richness increases, more sophisticated reasoning-oriented prompting strategies become advantageous. Practitioners should match their prompting approaches to the available API abstraction level: Basic Prompts for no abstraction, In-context Learning for low abstraction (A1), and Chain-of-Thought techniques for higher abstraction levels (A2-A4).

\paragraph{Vulnerability-Specific Approaches:} Different vulnerability types benefit from distinct abstraction levels. Memory boundary-related vulnerabilities can be effectively detected even with minimal API abstraction (A1), while resource management vulnerabilities require higher abstraction levels (A3-A4) for optimal detection. Security teams should consider tailoring their approach based on the vulnerability types of greatest concern.

\paragraph{Precision-Recall Trade-offs:} Our PacVD approach demonstrates that it's possible to maintain high recall without sacrificing precision, a critical requirement for practical deployment. Organizations should consider abstraction-based approaches to reduce false positives while maintaining detection sensitivity.

\paragraph{Continuous Learning Ecosystems:} Our findings suggest the potential for continuously improving vulnerability detection systems through feedback loops. As new vulnerabilities are discovered, they could be incorporated into the in-context examples provided to LLMs, enabling the system to adapt to emerging threat patterns without requiring full model retraining.

\subsection{Future Research Directions and Challenges}

\paragraph{Information Saturation Point}: Our findings suggest that there exists an information saturation point (\ie around A3 level) beyond which additional API details yield diminishing or negative returns. Future research should investigate this phenomenon across different domains to develop a more generalized theory of optimal information density for LLM reasoning.

\paragraph{Multi-Modal Vulnerability Detection:} Building on our API abstraction approach, future research could explore combining multiple information modalities, such as API abstractions, data flow graphs, and natural language documentation. Understanding how these different information types complement each other could lead to more comprehensive vulnerability detection capabilities.

\paragraph{Model Architecture Influences}: The varying sensitivity of different models to API abstraction levels suggests fundamental differences in how model architectures process structured information. Future work should investigate the architectural features that enable models like DeepSeek-R1 to maintain consistent performance across abstraction levels.

\paragraph{Abstraction Precision}: Our results validate the importance of path-sensitive abstractions for vulnerability detection. However, they also demonstrate that the relationship between abstraction precision and effectiveness is non-linear. This finding warrants deeper exploration of abstract interpretation theory in the context of LLM-based analysis.

\paragraph{Cross-Modal Information Integration}: The effectiveness of combining structured API information with natural language prompts suggests promising avenues for research on how LLMs integrate different information modalities. This could lead to novel approaches for software security beyond vulnerability detection.

\paragraph{API Abstraction for Education}: The clear performance improvements achieved through API abstraction could inform better approaches for teaching secure coding practices, helping developers understand the relationship between API usage patterns and potential vulnerabilities.

\paragraph{Programming Language Diversity:} While our study focused on C/C++ vulnerabilities, implementing similar API abstraction techniques across different programming languages presents unique challenges. Languages with dynamic typing (e.g., Python, JavaScript) would require runtime information to effectively abstract API behavior. In contrast, strongly-typed languages (e.g., Rust, Java) might benefit from leveraging type information to enhance abstraction quality. Future implementations would need language-specific analyzers to extract equivalent API abstractions while preserving the semantic information that makes our approach effective.


\paragraph{Model Deployment and Maintenance:} Implementing LLM-based vulnerability detection in production environments introduces challenges related to model versioning, updating, and drift monitoring. Organizations would need strategies to evaluate whether model performance deteriorates over time as new APIs and vulnerability patterns emerge, potentially requiring periodic retraining or fine-tuning.

\section{Related Work}

\subsection{Current Advances in Vulnerability Detection}

Recent advances in vulnerability detection leverage deep learning and large language models (LLMs) to analyze complex code patterns, categorized into traditional machine learning methods, sequence-based learning, graph-based models, and LLM-based approaches.

Traditional machine learning-based approaches rely on handcrafted features to identify vulnerabilities, but their limited adaptability and reliance on expert intervention hinder their effectiveness across diverse codebases \cite{du2019leopard,zimmermann2007predicting, li2019comparative}.
Sequence-based deep learning methods treat code as token sequences, significantly improving detection accuracy by capturing syntactic and semantic information, though they struggle with fully representing the hierarchical structure of code \cite{li2018vuldeepecker,li2021sysevr, wu2022code}.
Graph-based deep learning models enhance vulnerability detection by representing code as graphs, effectively capturing relationships and facilitating comprehensive structural analysis, thus improving accuracy and robustness \cite{furno2018graph,  zhou2019devign, yamaguchi2014modeling}.
LLM-Based Methods utilize pre-trained models for vulnerability detection, demonstrating adaptability and effectiveness through fine-tuning and prompt engineering, significantly transforming the field \cite{li2023llm,steenhoek2024comprehensive,zhang2024prompt, su2023optimizing, chen2024bridge, li2024llm}.
Recent advances in prompt-enhanced LLM methods highlight the potential of LLMs for identifying vulnerabilities through carefully designed prompts, improving detection capabilities, and addressing complex security challenges in code analysis \cite{purba2023software,zhou2024large,ding2024vulnerability,yang2024dlap}.
Our approach follows a similar strategy but incorporates more contextual information extracted by program analysis.




\subsection{Contextual Representations in Vulnerability Detection}

Recent studies in vulnerability detection highlight the importance of extracting contextual information from data flow, API calls, and control flow representations to enhance detection effectiveness.
Control flow and data flow analysis are essential for understanding program semantics \cite{cheng2019static, ghorbanzadeh2020anovul, steenhoek2024dataflow, jiang2024dfept}. Exemplified by the SySeVR framework, which uses control and data dependency graphs to improve vulnerability prediction \cite{li2021sysevr}.
Advancements in control flow analysis, like Zhou et al.'s~\cite{zhou2019method} improved the control flow graph method, enhancing traditional representations for more accurate vulnerability detection.
API call sequences and program dependence graphs (PDGs) play a critical role in vulnerability detection \cite{fadadu2020evading, shankarapani2011malware, li2021pdgraph, li2021vulnerability}; Zhou et al.'s~\cite{zhou2019devign} Devign model integrates ASTs, CFGs, and PDGs to capture complex code interdependencies, while Alon et al.'s\cite{alon2019code2vec} Code2Vec learns vector representations from AST paths to enhance understanding of API call sequences.
Building on these methodologies, Zhang et al. proposed a framework that combines context slicing with multi-feature integration, improving detection accuracy by focusing on relevant code segments \cite{zhang2024context}.
However, these approaches may still introduce a lot of noise that is irrelevant to the vulnerabilities.
Differently, \toolname focuses on the key features that are relevant to the vulnerability by conducting primitive API-guided abstraction.
Moreover, we specifically focus on studying how different abstraction levels affect the performance of different LLMs.

\section{Conclusion}
In this work, we address the challenges of vulnerability detection by leveraging primitive API abstractions and context-enhanced techniques with large language models (LLMs). 
Our proposed method, \toolname, combines API abstraction and prompt engineering to enhance LLM performance, demonstrating that different abstraction levels and tailored prompt strategies significantly improve detection capabilities. 
Experimental results showed substantial improvements over baselines, highlighting the value of combining appropriate abstraction levels with effective prompt strategies to enhance vulnerability detection. 
These insights pave the way for more robust and interpretable software security solutions.

\section{Data Availability}

We open-source our datasets and experimental details to facilitate future research, which is available at: \textbf{\textcolor{violet}{\url{https://github.com/DoeSEResearch/PacVD.git}}}

\bibliographystyle{ACM-Reference-Format}


\begin{thebibliography}{0}


\ifx \showCODEN    \undefined \def \showCODEN     #1{\unskip}     \fi
\ifx \showDOI      \undefined \def \showDOI       #1{#1}\fi
\ifx \showISBNx    \undefined \def \showISBNx     #1{\unskip}     \fi
\ifx \showISBNxiii \undefined \def \showISBNxiii  #1{\unskip}     \fi
\ifx \showISSN     \undefined \def \showISSN      #1{\unskip}     \fi
\ifx \showLCCN     \undefined \def \showLCCN      #1{\unskip}     \fi
\ifx \shownote     \undefined \def \shownote      #1{#1}          \fi
\ifx \showarticletitle \undefined \def \showarticletitle #1{#1}   \fi
\ifx \showURL      \undefined \def \showURL       {\relax}        \fi
\providecommand\bibfield[2]{#2}
\providecommand\bibinfo[2]{#2}
\providecommand\natexlab[1]{#1}
\providecommand\showeprint[2][]{arXiv:#2}

\end{thebibliography}


\begin{thebibliography}{65}


\ifx \showCODEN    \undefined \def \showCODEN     #1{\unskip}     \fi
\ifx \showDOI      \undefined \def \showDOI       #1{#1}\fi
\ifx \showISBNx    \undefined \def \showISBNx     #1{\unskip}     \fi
\ifx \showISBNxiii \undefined \def \showISBNxiii  #1{\unskip}     \fi
\ifx \showISSN     \undefined \def \showISSN      #1{\unskip}     \fi
\ifx \showLCCN     \undefined \def \showLCCN      #1{\unskip}     \fi
\ifx \shownote     \undefined \def \shownote      #1{#1}          \fi
\ifx \showarticletitle \undefined \def \showarticletitle #1{#1}   \fi
\ifx \showURL      \undefined \def \showURL       {\relax}        \fi
\providecommand\bibfield[2]{#2}
\providecommand\bibinfo[2]{#2}
\providecommand\natexlab[1]{#1}
\providecommand\showeprint[2][]{arXiv:#2}

\bibitem[\protect\citeauthoryear{Achiam, Adler, Agarwal, Ahmad, Akkaya, Aleman,
  Almeida, Altenschmidt, Altman, Anadkat, et~al\mbox{.}}{Achiam
  et~al\mbox{.}}{2023}]%
        {achiam2023gpt}
\bibfield{author}{\bibinfo{person}{Josh Achiam}, \bibinfo{person}{Steven
  Adler}, \bibinfo{person}{Sandhini Agarwal}, \bibinfo{person}{Lama Ahmad},
  \bibinfo{person}{Ilge Akkaya}, \bibinfo{person}{Florencia~Leoni Aleman},
  \bibinfo{person}{Diogo Almeida}, \bibinfo{person}{Janko Altenschmidt},
  \bibinfo{person}{Sam Altman}, \bibinfo{person}{Shyamal Anadkat},
  {et~al\mbox{.}}} \bibinfo{year}{2023}\natexlab{}.
\newblock \showarticletitle{Gpt-4 technical report}.
\newblock \bibinfo{journal}{\emph{arXiv preprint arXiv:2303.08774}}
  (\bibinfo{year}{2023}).
\newblock


\bibitem[\protect\citeauthoryear{Alon, Zilberstein, Levy, and Yahav}{Alon
  et~al\mbox{.}}{2019}]%
        {alon2019code2vec}
\bibfield{author}{\bibinfo{person}{Uri Alon}, \bibinfo{person}{Meital
  Zilberstein}, \bibinfo{person}{Omer Levy}, {and} \bibinfo{person}{Eran
  Yahav}.} \bibinfo{year}{2019}\natexlab{}.
\newblock \showarticletitle{code2vec: Learning distributed representations of
  code}.
\newblock \bibinfo{journal}{\emph{Proceedings of the ACM on Programming
  Languages}} \bibinfo{volume}{3}, \bibinfo{number}{POPL}
  (\bibinfo{year}{2019}), \bibinfo{pages}{1--29}.
\newblock


\bibitem[\protect\citeauthoryear{Brown}{Brown}{2020}]%
        {brown2020language}
\bibfield{author}{\bibinfo{person}{Tom~B Brown}.}
  \bibinfo{year}{2020}\natexlab{}.
\newblock \showarticletitle{Language models are few-shot learners}.
\newblock \bibinfo{journal}{\emph{arXiv preprint arXiv:2005.14165}}
  (\bibinfo{year}{2020}).
\newblock


\bibitem[\protect\citeauthoryear{Chakraborty, Krishna, Ding, and
  Ray}{Chakraborty et~al\mbox{.}}{2022}]%
        {chakraborty2021deep}
\bibfield{author}{\bibinfo{person}{Saikat Chakraborty}, \bibinfo{person}{Rahul
  Krishna}, \bibinfo{person}{Yangruibo Ding}, {and} \bibinfo{person}{Baishakhi
  Ray}.} \bibinfo{year}{2022}\natexlab{}.
\newblock \showarticletitle{Deep Learning Based Vulnerability Detection: Are We
  There Yet?}
\newblock \bibinfo{journal}{\emph{IEEE Transactions on Software Engineering}}
  \bibinfo{volume}{48}, \bibinfo{number}{9} (\bibinfo{year}{2022}),
  \bibinfo{pages}{3280--3296}.
\newblock
\urldef\tempurl%
\url{https://doi.org/10.1109/TSE.2021.3087402}
\showDOI{\tempurl}


\bibitem[\protect\citeauthoryear{Chen and Ng}{Chen and Ng}{2004}]%
        {chen2004marriage}
\bibfield{author}{\bibinfo{person}{Lei Chen} {and} \bibinfo{person}{Raymond
  Ng}.} \bibinfo{year}{2004}\natexlab{}.
\newblock \showarticletitle{On the marriage of lp-norms and edit distance}. In
  \bibinfo{booktitle}{\emph{Proceedings of the Thirtieth international
  conference on Very large data bases-Volume 30}}. \bibinfo{pages}{792--803}.
\newblock


\bibitem[\protect\citeauthoryear{Chen, Gao, Yang, Zhang, and Liao}{Chen
  et~al\mbox{.}}{2024}]%
        {chen2024bridge}
\bibfield{author}{\bibinfo{person}{Yujia Chen}, \bibinfo{person}{Cuiyun Gao},
  \bibinfo{person}{Zezhou Yang}, \bibinfo{person}{Hongyu Zhang}, {and}
  \bibinfo{person}{Qing Liao}.} \bibinfo{year}{2024}\natexlab{}.
\newblock \showarticletitle{Bridge and Hint: Extending Pre-trained Language
  Models for Long-Range Code}. In \bibinfo{booktitle}{\emph{Proceedings of the
  33rd ACM SIGSOFT International Symposium on Software Testing and Analysis}}.
  \bibinfo{pages}{274--286}.
\newblock


\bibitem[\protect\citeauthoryear{Chen, Wang, Yan, Sui, and Xue}{Chen
  et~al\mbox{.}}{2021}]%
        {10.1145/3460319.3464807}
\bibfield{author}{\bibinfo{person}{Zhe Chen}, \bibinfo{person}{Chong Wang},
  \bibinfo{person}{Junqi Yan}, \bibinfo{person}{Yulei Sui}, {and}
  \bibinfo{person}{Jingling Xue}.} \bibinfo{year}{2021}\natexlab{}.
\newblock \showarticletitle{Runtime detection of memory errors with smart
  status}. In \bibinfo{booktitle}{\emph{Proceedings of the 30th ACM SIGSOFT
  International Symposium on Software Testing and Analysis}} (Virtual, Denmark)
  \emph{(\bibinfo{series}{ISSTA 2021})}. \bibinfo{publisher}{Association for
  Computing Machinery}, \bibinfo{address}{New York, NY, USA},
  \bibinfo{pages}{296–308}.
\newblock
\showISBNx{9781450384599}
\urldef\tempurl%
\url{https://doi.org/10.1145/3460319.3464807}
\showDOI{\tempurl}


\bibitem[\protect\citeauthoryear{Cheng, Wang, Hua, Zhang, Xu, Yi, and
  Sui}{Cheng et~al\mbox{.}}{2019}]%
        {cheng2019static}
\bibfield{author}{\bibinfo{person}{Xiao Cheng}, \bibinfo{person}{Haoyu Wang},
  \bibinfo{person}{Jiayi Hua}, \bibinfo{person}{Miao Zhang},
  \bibinfo{person}{Guoai Xu}, \bibinfo{person}{Li Yi}, {and}
  \bibinfo{person}{Yulei Sui}.} \bibinfo{year}{2019}\natexlab{}.
\newblock \showarticletitle{Static detection of control-flow-related
  vulnerabilities using graph embedding}. In \bibinfo{booktitle}{\emph{2019
  24th International Conference on Engineering of Complex Computer Systems
  (ICECCS)}}. IEEE, \bibinfo{pages}{41--50}.
\newblock


\bibitem[\protect\citeauthoryear{{CVE}}{{CVE}}{2021}]%
        {CVE}
\bibfield{author}{\bibinfo{person}{{CVE}}.} \bibinfo{year}{2021}\natexlab{}.
\newblock \bibinfo{title}{NVD}.
\newblock
\newblock
\newblock
\shownote{\url{https://cve.mitre.org/}.}


\bibitem[\protect\citeauthoryear{DeepSeek-AI, Guo, Yang, Zhang, Song, Zhang,
  Xu, Zhu, Ma, Wang, Bi, Zhang, Yu, Wu, Wu, Gou, Shao, Li, Gao, Liu, Xue, Wang,
  Wu, Feng, Lu, Zhao, Deng, Zhang, Ruan, Dai, Chen, Ji, Li, Lin, Dai, Luo, Hao,
  Chen, Li, Zhang, Bao, Xu, Wang, Ding, Xin, Gao, Qu, Li, Guo, Li, Wang, Chen,
  Yuan, Qiu, Li, Cai, Ni, Liang, Chen, Dong, Hu, Gao, Guan, Huang, Yu, Wang,
  Zhang, Zhao, Wang, Zhang, Xu, Xia, Zhang, Zhang, Tang, Li, Wang, Li, Tian,
  Huang, Zhang, Wang, Chen, Du, Ge, Zhang, Pan, Wang, Chen, Jin, Chen, Lu,
  Zhou, Chen, Ye, Wang, Yu, Zhou, Pan, Li, Zhou, Wu, Ye, Yun, Pei, Sun, Wang,
  Zeng, Zhao, Liu, Liang, Gao, Yu, Zhang, Xiao, An, Liu, Wang, Chen, Nie,
  Cheng, Liu, Xie, Liu, Yang, Li, Su, Lin, Li, Jin, Shen, Chen, Sun, Wang,
  Song, Zhou, Wang, Shan, Li, Wang, Wei, Zhang, Xu, Li, Zhao, Sun, Wang, Yu,
  Zhang, Shi, Xiong, He, Piao, Wang, Tan, Ma, Liu, Guo, Ou, Wang, Gong, Zou,
  He, Xiong, Luo, You, Liu, Zhou, Zhu, Xu, Huang, Li, Zheng, Zhu, Ma, Tang,
  Zha, Yan, Ren, Ren, Sha, Fu, Xu, Xie, Zhang, Hao, Ma, Yan, Wu, Gu, Zhu, Liu,
  Li, Xie, Song, Pan, Huang, Xu, Zhang, and Zhang}{DeepSeek-AI
  et~al\mbox{.}}{2025}]%
        {deepseekai2025deepseekr1incentivizingreasoningcapability}
\bibfield{author}{\bibinfo{person}{DeepSeek-AI}, \bibinfo{person}{Daya Guo},
  \bibinfo{person}{Dejian Yang}, \bibinfo{person}{Haowei Zhang},
  \bibinfo{person}{Junxiao Song}, \bibinfo{person}{Ruoyu Zhang},
  \bibinfo{person}{Runxin Xu}, \bibinfo{person}{Qihao Zhu},
  \bibinfo{person}{Shirong Ma}, \bibinfo{person}{Peiyi Wang},
  \bibinfo{person}{Xiao Bi}, \bibinfo{person}{Xiaokang Zhang},
  \bibinfo{person}{Xingkai Yu}, \bibinfo{person}{Yu Wu}, \bibinfo{person}{Z.~F.
  Wu}, \bibinfo{person}{Zhibin Gou}, \bibinfo{person}{Zhihong Shao},
  \bibinfo{person}{Zhuoshu Li}, \bibinfo{person}{Ziyi Gao},
  \bibinfo{person}{Aixin Liu}, \bibinfo{person}{Bing Xue},
  \bibinfo{person}{Bingxuan Wang}, \bibinfo{person}{Bochao Wu},
  \bibinfo{person}{Bei Feng}, \bibinfo{person}{Chengda Lu},
  \bibinfo{person}{Chenggang Zhao}, \bibinfo{person}{Chengqi Deng},
  \bibinfo{person}{Chenyu Zhang}, \bibinfo{person}{Chong Ruan},
  \bibinfo{person}{Damai Dai}, \bibinfo{person}{Deli Chen},
  \bibinfo{person}{Dongjie Ji}, \bibinfo{person}{Erhang Li},
  \bibinfo{person}{Fangyun Lin}, \bibinfo{person}{Fucong Dai},
  \bibinfo{person}{Fuli Luo}, \bibinfo{person}{Guangbo Hao},
  \bibinfo{person}{Guanting Chen}, \bibinfo{person}{Guowei Li},
  \bibinfo{person}{H. Zhang}, \bibinfo{person}{Han Bao},
  \bibinfo{person}{Hanwei Xu}, \bibinfo{person}{Haocheng Wang},
  \bibinfo{person}{Honghui Ding}, \bibinfo{person}{Huajian Xin},
  \bibinfo{person}{Huazuo Gao}, \bibinfo{person}{Hui Qu}, \bibinfo{person}{Hui
  Li}, \bibinfo{person}{Jianzhong Guo}, \bibinfo{person}{Jiashi Li},
  \bibinfo{person}{Jiawei Wang}, \bibinfo{person}{Jingchang Chen},
  \bibinfo{person}{Jingyang Yuan}, \bibinfo{person}{Junjie Qiu},
  \bibinfo{person}{Junlong Li}, \bibinfo{person}{J.~L. Cai},
  \bibinfo{person}{Jiaqi Ni}, \bibinfo{person}{Jian Liang},
  \bibinfo{person}{Jin Chen}, \bibinfo{person}{Kai Dong}, \bibinfo{person}{Kai
  Hu}, \bibinfo{person}{Kaige Gao}, \bibinfo{person}{Kang Guan},
  \bibinfo{person}{Kexin Huang}, \bibinfo{person}{Kuai Yu},
  \bibinfo{person}{Lean Wang}, \bibinfo{person}{Lecong Zhang},
  \bibinfo{person}{Liang Zhao}, \bibinfo{person}{Litong Wang},
  \bibinfo{person}{Liyue Zhang}, \bibinfo{person}{Lei Xu},
  \bibinfo{person}{Leyi Xia}, \bibinfo{person}{Mingchuan Zhang},
  \bibinfo{person}{Minghua Zhang}, \bibinfo{person}{Minghui Tang},
  \bibinfo{person}{Meng Li}, \bibinfo{person}{Miaojun Wang},
  \bibinfo{person}{Mingming Li}, \bibinfo{person}{Ning Tian},
  \bibinfo{person}{Panpan Huang}, \bibinfo{person}{Peng Zhang},
  \bibinfo{person}{Qiancheng Wang}, \bibinfo{person}{Qinyu Chen},
  \bibinfo{person}{Qiushi Du}, \bibinfo{person}{Ruiqi Ge},
  \bibinfo{person}{Ruisong Zhang}, \bibinfo{person}{Ruizhe Pan},
  \bibinfo{person}{Runji Wang}, \bibinfo{person}{R.~J. Chen},
  \bibinfo{person}{R.~L. Jin}, \bibinfo{person}{Ruyi Chen},
  \bibinfo{person}{Shanghao Lu}, \bibinfo{person}{Shangyan Zhou},
  \bibinfo{person}{Shanhuang Chen}, \bibinfo{person}{Shengfeng Ye},
  \bibinfo{person}{Shiyu Wang}, \bibinfo{person}{Shuiping Yu},
  \bibinfo{person}{Shunfeng Zhou}, \bibinfo{person}{Shuting Pan},
  \bibinfo{person}{S.~S. Li}, \bibinfo{person}{Shuang Zhou},
  \bibinfo{person}{Shaoqing Wu}, \bibinfo{person}{Shengfeng Ye},
  \bibinfo{person}{Tao Yun}, \bibinfo{person}{Tian Pei},
  \bibinfo{person}{Tianyu Sun}, \bibinfo{person}{T. Wang},
  \bibinfo{person}{Wangding Zeng}, \bibinfo{person}{Wanjia Zhao},
  \bibinfo{person}{Wen Liu}, \bibinfo{person}{Wenfeng Liang},
  \bibinfo{person}{Wenjun Gao}, \bibinfo{person}{Wenqin Yu},
  \bibinfo{person}{Wentao Zhang}, \bibinfo{person}{W.~L. Xiao},
  \bibinfo{person}{Wei An}, \bibinfo{person}{Xiaodong Liu},
  \bibinfo{person}{Xiaohan Wang}, \bibinfo{person}{Xiaokang Chen},
  \bibinfo{person}{Xiaotao Nie}, \bibinfo{person}{Xin Cheng},
  \bibinfo{person}{Xin Liu}, \bibinfo{person}{Xin Xie},
  \bibinfo{person}{Xingchao Liu}, \bibinfo{person}{Xinyu Yang},
  \bibinfo{person}{Xinyuan Li}, \bibinfo{person}{Xuecheng Su},
  \bibinfo{person}{Xuheng Lin}, \bibinfo{person}{X.~Q. Li},
  \bibinfo{person}{Xiangyue Jin}, \bibinfo{person}{Xiaojin Shen},
  \bibinfo{person}{Xiaosha Chen}, \bibinfo{person}{Xiaowen Sun},
  \bibinfo{person}{Xiaoxiang Wang}, \bibinfo{person}{Xinnan Song},
  \bibinfo{person}{Xinyi Zhou}, \bibinfo{person}{Xianzu Wang},
  \bibinfo{person}{Xinxia Shan}, \bibinfo{person}{Y.~K. Li},
  \bibinfo{person}{Y.~Q. Wang}, \bibinfo{person}{Y.~X. Wei},
  \bibinfo{person}{Yang Zhang}, \bibinfo{person}{Yanhong Xu},
  \bibinfo{person}{Yao Li}, \bibinfo{person}{Yao Zhao},
  \bibinfo{person}{Yaofeng Sun}, \bibinfo{person}{Yaohui Wang},
  \bibinfo{person}{Yi Yu}, \bibinfo{person}{Yichao Zhang},
  \bibinfo{person}{Yifan Shi}, \bibinfo{person}{Yiliang Xiong},
  \bibinfo{person}{Ying He}, \bibinfo{person}{Yishi Piao},
  \bibinfo{person}{Yisong Wang}, \bibinfo{person}{Yixuan Tan},
  \bibinfo{person}{Yiyang Ma}, \bibinfo{person}{Yiyuan Liu},
  \bibinfo{person}{Yongqiang Guo}, \bibinfo{person}{Yuan Ou},
  \bibinfo{person}{Yuduan Wang}, \bibinfo{person}{Yue Gong},
  \bibinfo{person}{Yuheng Zou}, \bibinfo{person}{Yujia He},
  \bibinfo{person}{Yunfan Xiong}, \bibinfo{person}{Yuxiang Luo},
  \bibinfo{person}{Yuxiang You}, \bibinfo{person}{Yuxuan Liu},
  \bibinfo{person}{Yuyang Zhou}, \bibinfo{person}{Y.~X. Zhu},
  \bibinfo{person}{Yanhong Xu}, \bibinfo{person}{Yanping Huang},
  \bibinfo{person}{Yaohui Li}, \bibinfo{person}{Yi Zheng},
  \bibinfo{person}{Yuchen Zhu}, \bibinfo{person}{Yunxian Ma},
  \bibinfo{person}{Ying Tang}, \bibinfo{person}{Yukun Zha},
  \bibinfo{person}{Yuting Yan}, \bibinfo{person}{Z.~Z. Ren},
  \bibinfo{person}{Zehui Ren}, \bibinfo{person}{Zhangli Sha},
  \bibinfo{person}{Zhe Fu}, \bibinfo{person}{Zhean Xu}, \bibinfo{person}{Zhenda
  Xie}, \bibinfo{person}{Zhengyan Zhang}, \bibinfo{person}{Zhewen Hao},
  \bibinfo{person}{Zhicheng Ma}, \bibinfo{person}{Zhigang Yan},
  \bibinfo{person}{Zhiyu Wu}, \bibinfo{person}{Zihui Gu},
  \bibinfo{person}{Zijia Zhu}, \bibinfo{person}{Zijun Liu},
  \bibinfo{person}{Zilin Li}, \bibinfo{person}{Ziwei Xie},
  \bibinfo{person}{Ziyang Song}, \bibinfo{person}{Zizheng Pan},
  \bibinfo{person}{Zhen Huang}, \bibinfo{person}{Zhipeng Xu},
  \bibinfo{person}{Zhongyu Zhang}, {and} \bibinfo{person}{Zhen Zhang}.}
  \bibinfo{year}{2025}\natexlab{}.
\newblock \bibinfo{title}{DeepSeek-R1: Incentivizing Reasoning Capability in
  LLMs via Reinforcement Learning}.
\newblock
\newblock
\showeprint[arxiv]{2501.12948}~[cs.CL]
\urldef\tempurl%
\url{https://arxiv.org/abs/2501.12948}
\showURL{%
\tempurl}


\bibitem[\protect\citeauthoryear{Ding, Fu, Ibrahim, Sitawarin, Chen, Alomair,
  Wagner, Ray, and Chen}{Ding et~al\mbox{.}}{2024}]%
        {ding2024vulnerability}
\bibfield{author}{\bibinfo{person}{Yangruibo Ding}, \bibinfo{person}{Yanjun
  Fu}, \bibinfo{person}{Omniyyah Ibrahim}, \bibinfo{person}{Chawin Sitawarin},
  \bibinfo{person}{Xinyun Chen}, \bibinfo{person}{Basel Alomair},
  \bibinfo{person}{David Wagner}, \bibinfo{person}{Baishakhi Ray}, {and}
  \bibinfo{person}{Yizheng Chen}.} \bibinfo{year}{2024}\natexlab{}.
\newblock \showarticletitle{Vulnerability detection with code language models:
  How far are we?}
\newblock \bibinfo{journal}{\emph{arXiv preprint arXiv:2403.18624}}
  (\bibinfo{year}{2024}).
\newblock


\bibitem[\protect\citeauthoryear{Du, Chen, Li, Guo, Zhou, Liu, and Jiang}{Du
  et~al\mbox{.}}{2019}]%
        {du2019leopard}
\bibfield{author}{\bibinfo{person}{Xiaoning Du}, \bibinfo{person}{Bihuan Chen},
  \bibinfo{person}{Yuekang Li}, \bibinfo{person}{Jianmin Guo},
  \bibinfo{person}{Yaqin Zhou}, \bibinfo{person}{Yang Liu}, {and}
  \bibinfo{person}{Yu Jiang}.} \bibinfo{year}{2019}\natexlab{}.
\newblock \showarticletitle{Leopard: Identifying vulnerable code for
  vulnerability assessment through program metrics}. In
  \bibinfo{booktitle}{\emph{2019 IEEE/ACM 41st International Conference on
  Software Engineering (ICSE)}}. IEEE, \bibinfo{pages}{60--71}.
\newblock


\bibitem[\protect\citeauthoryear{Du, Wen, Zhu, Xie, Ji, Liu, Shi, and Jin}{Du
  et~al\mbox{.}}{2024a}]%
        {du-etal-2024-generalization}
\bibfield{author}{\bibinfo{person}{Xiaohu Du}, \bibinfo{person}{Ming Wen},
  \bibinfo{person}{Jiahao Zhu}, \bibinfo{person}{Zifan Xie},
  \bibinfo{person}{Bin Ji}, \bibinfo{person}{Huijun Liu},
  \bibinfo{person}{Xuanhua Shi}, {and} \bibinfo{person}{Hai Jin}.}
  \bibinfo{year}{2024}\natexlab{a}.
\newblock \showarticletitle{Generalization-Enhanced Code Vulnerability
  Detection via Multi-Task Instruction Fine-Tuning}. In
  \bibinfo{booktitle}{\emph{Findings of the Association for Computational
  Linguistics: ACL 2024}}, \bibfield{editor}{\bibinfo{person}{Lun-Wei Ku},
  \bibinfo{person}{Andre Martins}, {and} \bibinfo{person}{Vivek Srikumar}}
  (Eds.). \bibinfo{publisher}{Association for Computational Linguistics},
  \bibinfo{address}{Bangkok, Thailand}, \bibinfo{pages}{10507--10521}.
\newblock
\urldef\tempurl%
\url{https://doi.org/10.18653/v1/2024.findings-acl.625}
\showDOI{\tempurl}


\bibitem[\protect\citeauthoryear{Du, Zheng, Wang, Feng, Deng, Liu, Chen, Peng,
  Ma, and Lou}{Du et~al\mbox{.}}{2024b}]%
        {du2024vulragenhancingllmbasedvulnerability}
\bibfield{author}{\bibinfo{person}{Xueying Du}, \bibinfo{person}{Geng Zheng},
  \bibinfo{person}{Kaixin Wang}, \bibinfo{person}{Jiayi Feng},
  \bibinfo{person}{Wentai Deng}, \bibinfo{person}{Mingwei Liu},
  \bibinfo{person}{Bihuan Chen}, \bibinfo{person}{Xin Peng},
  \bibinfo{person}{Tao Ma}, {and} \bibinfo{person}{Yiling Lou}.}
  \bibinfo{year}{2024}\natexlab{b}.
\newblock \bibinfo{title}{Vul-RAG: Enhancing LLM-based Vulnerability Detection
  via Knowledge-level RAG}.
\newblock
\newblock
\showeprint[arxiv]{2406.11147}~[cs.SE]
\urldef\tempurl%
\url{https://arxiv.org/abs/2406.11147}
\showURL{%
\tempurl}


\bibitem[\protect\citeauthoryear{Fadadu, Handa, Kumar, and Shukla}{Fadadu
  et~al\mbox{.}}{2020}]%
        {fadadu2020evading}
\bibfield{author}{\bibinfo{person}{Fenil Fadadu}, \bibinfo{person}{Anand
  Handa}, \bibinfo{person}{Nitesh Kumar}, {and} \bibinfo{person}{Sandeep~Kumar
  Shukla}.} \bibinfo{year}{2020}\natexlab{}.
\newblock \showarticletitle{Evading API call sequence based malware
  classifiers}. In \bibinfo{booktitle}{\emph{Information and Communications
  Security: 21st International Conference, ICICS 2019, Beijing, China, December
  15--17, 2019, Revised Selected Papers 21}}. Springer,
  \bibinfo{pages}{18--33}.
\newblock


\bibitem[\protect\citeauthoryear{Furno, El~Faouzi, Sharma, Cammarota, and
  Zimeo}{Furno et~al\mbox{.}}{2018}]%
        {furno2018graph}
\bibfield{author}{\bibinfo{person}{Angelo Furno}, \bibinfo{person}{Nour-Eddin
  El~Faouzi}, \bibinfo{person}{Rajesh Sharma}, \bibinfo{person}{Valerio
  Cammarota}, {and} \bibinfo{person}{Eugenio Zimeo}.}
  \bibinfo{year}{2018}\natexlab{}.
\newblock \showarticletitle{A graph-based framework for real-time vulnerability
  assessment of road networks}. In \bibinfo{booktitle}{\emph{2018 IEEE
  International Conference on Smart Computing (SMARTCOMP)}}. IEEE,
  \bibinfo{pages}{234--241}.
\newblock


\bibitem[\protect\citeauthoryear{Ghorbanzadeh and Reza~Shahriari}{Ghorbanzadeh
  and Reza~Shahriari}{2020}]%
        {ghorbanzadeh2020anovul}
\bibfield{author}{\bibinfo{person}{Mahmoud Ghorbanzadeh} {and}
  \bibinfo{person}{Hamid Reza~Shahriari}.} \bibinfo{year}{2020}\natexlab{}.
\newblock \showarticletitle{ANOVUL: Detection of logic vulnerabilities in
  annotated programs via data and control flow analysis}.
\newblock \bibinfo{journal}{\emph{IET Information Security}}
  \bibinfo{volume}{14}, \bibinfo{number}{3} (\bibinfo{year}{2020}),
  \bibinfo{pages}{352--364}.
\newblock


\bibitem[\protect\citeauthoryear{Han, Gao, Liu, Zhang, and Zhang}{Han
  et~al\mbox{.}}{2024}]%
        {han2024parameterefficientfinetuninglargemodels}
\bibfield{author}{\bibinfo{person}{Zeyu Han}, \bibinfo{person}{Chao Gao},
  \bibinfo{person}{Jinyang Liu}, \bibinfo{person}{Jeff Zhang}, {and}
  \bibinfo{person}{Sai~Qian Zhang}.} \bibinfo{year}{2024}\natexlab{}.
\newblock \bibinfo{title}{Parameter-Efficient Fine-Tuning for Large Models: A
  Comprehensive Survey}.
\newblock
\newblock
\showeprint[arxiv]{2403.14608}~[cs.LG]
\urldef\tempurl%
\url{https://arxiv.org/abs/2403.14608}
\showURL{%
\tempurl}


\bibitem[\protect\citeauthoryear{Jiang, Sun, Gu, Wu, Wen, Hu, and Yan}{Jiang
  et~al\mbox{.}}{2024}]%
        {jiang2024dfept}
\bibfield{author}{\bibinfo{person}{Zhonghao Jiang}, \bibinfo{person}{Weifeng
  Sun}, \bibinfo{person}{Xiaoyan Gu}, \bibinfo{person}{Jiaxin Wu},
  \bibinfo{person}{Tao Wen}, \bibinfo{person}{Haibo Hu}, {and}
  \bibinfo{person}{Meng Yan}.} \bibinfo{year}{2024}\natexlab{}.
\newblock \showarticletitle{DFEPT: Data Flow Embedding for Enhancing
  Pre-Trained Model Based Vulnerability Detection}. In
  \bibinfo{booktitle}{\emph{Proceedings of the 15th Asia-Pacific Symposium on
  Internetware}}. \bibinfo{pages}{95--104}.
\newblock


\bibitem[\protect\citeauthoryear{Li and Shan}{Li and Shan}{2023}]%
        {li2023llm}
\bibfield{author}{\bibinfo{person}{Hongping Li} {and} \bibinfo{person}{Li
  Shan}.} \bibinfo{year}{2023}\natexlab{}.
\newblock \showarticletitle{LLM-based Vulnerability Detection}. In
  \bibinfo{booktitle}{\emph{2023 International Conference on Human-Centered
  Cognitive Systems (HCCS)}}. IEEE, \bibinfo{pages}{1--4}.
\newblock


\bibitem[\protect\citeauthoryear{Li, Song, Tan, Wang, and Liu}{Li
  et~al\mbox{.}}{2021a}]%
        {li2021pdgraph}
\bibfield{author}{\bibinfo{person}{Qiang Li}, \bibinfo{person}{Jinke Song},
  \bibinfo{person}{Dawei Tan}, \bibinfo{person}{Haining Wang}, {and}
  \bibinfo{person}{Jiqiang Liu}.} \bibinfo{year}{2021}\natexlab{a}.
\newblock \showarticletitle{Pdgraph: a large-scale empirical study on project
  dependency of security vulnerabilities}. In \bibinfo{booktitle}{\emph{2021
  51st Annual IEEE/IFIP International Conference on Dependable Systems and
  Networks (DSN)}}. IEEE, \bibinfo{pages}{161--173}.
\newblock


\bibitem[\protect\citeauthoryear{Li, Wang, and Nguyen}{Li
  et~al\mbox{.}}{2021b}]%
        {li2021vulnerability}
\bibfield{author}{\bibinfo{person}{Yi Li}, \bibinfo{person}{Shaohua Wang},
  {and} \bibinfo{person}{Tien~N Nguyen}.} \bibinfo{year}{2021}\natexlab{b}.
\newblock \showarticletitle{Vulnerability detection with fine-grained
  interpretations}. In \bibinfo{booktitle}{\emph{Proceedings of the 29th ACM
  Joint Meeting on European Software Engineering Conference and Symposium on
  the Foundations of Software Engineering}}. \bibinfo{pages}{292--303}.
\newblock


\bibitem[\protect\citeauthoryear{Li, Dutta, and Naik}{Li
  et~al\mbox{.}}{2024a}]%
        {li2024llm}
\bibfield{author}{\bibinfo{person}{Ziyang Li}, \bibinfo{person}{Saikat Dutta},
  {and} \bibinfo{person}{Mayur Naik}.} \bibinfo{year}{2024}\natexlab{a}.
\newblock \showarticletitle{LLM-Assisted Static Analysis for Detecting Security
  Vulnerabilities}.
\newblock \bibinfo{journal}{\emph{arXiv preprint arXiv:2405.17238}}
  (\bibinfo{year}{2024}).
\newblock


\bibitem[\protect\citeauthoryear{Li, Dutta, and Naik}{Li et~al\mbox{.}}{2025}]%
        {li2025irisllmassistedstaticanalysis}
\bibfield{author}{\bibinfo{person}{Ziyang Li}, \bibinfo{person}{Saikat Dutta},
  {and} \bibinfo{person}{Mayur Naik}.} \bibinfo{year}{2025}\natexlab{}.
\newblock \bibinfo{title}{IRIS: LLM-Assisted Static Analysis for Detecting
  Security Vulnerabilities}.
\newblock
\newblock
\showeprint[arxiv]{2405.17238}~[cs.CR]
\urldef\tempurl%
\url{https://arxiv.org/abs/2405.17238}
\showURL{%
\tempurl}


\bibitem[\protect\citeauthoryear{Li, Wang, Zou, Li, Zhang, Xu, Zhang, and
  Jin}{Li et~al\mbox{.}}{2024b}]%
        {li2024effectiveness}
\bibfield{author}{\bibinfo{person}{Zhen Li}, \bibinfo{person}{Ning Wang},
  \bibinfo{person}{Deqing Zou}, \bibinfo{person}{Yating Li},
  \bibinfo{person}{Ruqian Zhang}, \bibinfo{person}{Shouhuai Xu},
  \bibinfo{person}{Chao Zhang}, {and} \bibinfo{person}{Hai Jin}.}
  \bibinfo{year}{2024}\natexlab{b}.
\newblock \showarticletitle{On the Effectiveness of Function-Level
  Vulnerability Detectors for Inter-Procedural Vulnerabilities}. In
  \bibinfo{booktitle}{\emph{Proceedings of the IEEE/ACM 46th International
  Conference on Software Engineering}}. \bibinfo{pages}{1--12}.
\newblock


\bibitem[\protect\citeauthoryear{Li, Zou, Tang, Zhang, Sun, and Jin}{Li
  et~al\mbox{.}}{2019}]%
        {li2019comparative}
\bibfield{author}{\bibinfo{person}{Zhen Li}, \bibinfo{person}{Deqing Zou},
  \bibinfo{person}{Jing Tang}, \bibinfo{person}{Zhihao Zhang},
  \bibinfo{person}{Mingqian Sun}, {and} \bibinfo{person}{Hai Jin}.}
  \bibinfo{year}{2019}\natexlab{}.
\newblock \showarticletitle{A comparative study of deep learning-based
  vulnerability detection system}.
\newblock \bibinfo{journal}{\emph{IEEE Access}}  \bibinfo{volume}{7}
  (\bibinfo{year}{2019}), \bibinfo{pages}{103184--103197}.
\newblock


\bibitem[\protect\citeauthoryear{Li, Zou, Xu, Jin, Zhu, and Chen}{Li
  et~al\mbox{.}}{2021c}]%
        {li2021sysevr}
\bibfield{author}{\bibinfo{person}{Zhen Li}, \bibinfo{person}{Deqing Zou},
  \bibinfo{person}{Shouhuai Xu}, \bibinfo{person}{Hai Jin},
  \bibinfo{person}{Yawei Zhu}, {and} \bibinfo{person}{Zhaoxuan Chen}.}
  \bibinfo{year}{2021}\natexlab{c}.
\newblock \showarticletitle{Sysevr: A framework for using deep learning to
  detect software vulnerabilities}.
\newblock \bibinfo{journal}{\emph{IEEE Transactions on Dependable and Secure
  Computing}} \bibinfo{volume}{19}, \bibinfo{number}{4} (\bibinfo{year}{2021}),
  \bibinfo{pages}{2244--2258}.
\newblock


\bibitem[\protect\citeauthoryear{Li, Zou, Xu, Jin, Zhu, and Chen}{Li
  et~al\mbox{.}}{2022}]%
        {li2022sysevr}
\bibfield{author}{\bibinfo{person}{Zhen Li}, \bibinfo{person}{Deqing Zou},
  \bibinfo{person}{Shouhuai Xu}, \bibinfo{person}{Hai Jin},
  \bibinfo{person}{Yawei Zhu}, {and} \bibinfo{person}{Zhaoxuan Chen}.}
  \bibinfo{year}{2022}\natexlab{}.
\newblock \showarticletitle{SySeVR: A Framework for Using Deep Learning to
  Detect Software Vulnerabilities}.
\newblock \bibinfo{journal}{\emph{IEEE Transactions on Dependable and Secure
  Computing}} \bibinfo{volume}{19}, \bibinfo{number}{4} (\bibinfo{year}{2022}),
  \bibinfo{pages}{2244--2258}.
\newblock
\urldef\tempurl%
\url{https://doi.org/10.1109/TDSC.2021.3051525}
\showDOI{\tempurl}


\bibitem[\protect\citeauthoryear{Li, Zou, Xu, Ou, Jin, Wang, Deng, and
  Zhong}{Li et~al\mbox{.}}{2018}]%
        {li2018vuldeepecker}
\bibfield{author}{\bibinfo{person}{Zhen Li}, \bibinfo{person}{Deqing Zou},
  \bibinfo{person}{Shouhuai Xu}, \bibinfo{person}{Xinyu Ou},
  \bibinfo{person}{Hai Jin}, \bibinfo{person}{Sujuan Wang},
  \bibinfo{person}{Zhijun Deng}, {and} \bibinfo{person}{Yuyi Zhong}.}
  \bibinfo{year}{2018}\natexlab{}.
\newblock \showarticletitle{Vuldeepecker: A deep learning-based system for
  vulnerability detection}.
\newblock \bibinfo{journal}{\emph{arXiv preprint arXiv:1801.01681}}
  (\bibinfo{year}{2018}).
\newblock


\bibitem[\protect\citeauthoryear{Mao, Li, Hu, Liu, Xia, and Sun}{Mao
  et~al\mbox{.}}{2024}]%
        {Mao2024Towards}
\bibfield{author}{\bibinfo{person}{Qiheng Mao}, \bibinfo{person}{Zhenhao Li},
  \bibinfo{person}{Xing Hu}, \bibinfo{person}{Kui Liu}, \bibinfo{person}{Xin
  Xia}, {and} \bibinfo{person}{Jianling Sun}.} \bibinfo{year}{2024}\natexlab{}.
\newblock \bibinfo{title}{Towards Effectively Detecting and Explaining
  Vulnerabilities Using Large Language Models}.
\newblock
\newblock
\urldef\tempurl%
\url{https://doi.org/10.48550/arXiv.2406.09701}
\showDOI{\tempurl}


\bibitem[\protect\citeauthoryear{{National Vulnerability Database}}{{National
  Vulnerability Database}}{2021}]%
        {NVD}
\bibfield{author}{\bibinfo{person}{{National Vulnerability Database}}.}
  \bibinfo{year}{2021}\natexlab{}.
\newblock \bibinfo{title}{NVD}.
\newblock
\newblock
\newblock
\shownote{\url{https://nvd.nist.gov/}.}


\bibitem[\protect\citeauthoryear{Purba, Ghosh, Radford, and Chu}{Purba
  et~al\mbox{.}}{2023}]%
        {purba2023software}
\bibfield{author}{\bibinfo{person}{Moumita~Das Purba}, \bibinfo{person}{Arpita
  Ghosh}, \bibinfo{person}{Benjamin~J Radford}, {and} \bibinfo{person}{Bill
  Chu}.} \bibinfo{year}{2023}\natexlab{}.
\newblock \showarticletitle{Software vulnerability detection using large
  language models}. In \bibinfo{booktitle}{\emph{2023 IEEE 34th International
  Symposium on Software Reliability Engineering Workshops (ISSREW)}}. IEEE,
  \bibinfo{pages}{112--119}.
\newblock


\bibitem[\protect\citeauthoryear{Rinartha and Suryasa}{Rinartha and
  Suryasa}{2017}]%
        {rinartha2017comparative}
\bibfield{author}{\bibinfo{person}{Komang Rinartha} {and}
  \bibinfo{person}{Wayan Suryasa}.} \bibinfo{year}{2017}\natexlab{}.
\newblock \showarticletitle{Comparative study for better result on query
  suggestion of article searching with MySQL pattern matching and Jaccard
  similarity}. In \bibinfo{booktitle}{\emph{2017 5th International Conference
  on Cyber and IT Service Management (CITSM)}}. IEEE, \bibinfo{pages}{1--4}.
\newblock


\bibitem[\protect\citeauthoryear{Robertson, Zaragoza, et~al\mbox{.}}{Robertson
  et~al\mbox{.}}{2009}]%
        {robertson2009probabilistic}
\bibfield{author}{\bibinfo{person}{Stephen Robertson}, \bibinfo{person}{Hugo
  Zaragoza}, {et~al\mbox{.}}} \bibinfo{year}{2009}\natexlab{}.
\newblock \showarticletitle{The probabilistic relevance framework: BM25 and
  beyond}.
\newblock \bibinfo{journal}{\emph{Foundations and Trends{\textregistered} in
  Information Retrieval}} \bibinfo{volume}{3}, \bibinfo{number}{4}
  (\bibinfo{year}{2009}), \bibinfo{pages}{333--389}.
\newblock


\bibitem[\protect\citeauthoryear{Shanahan, McDonell, and Reynolds}{Shanahan
  et~al\mbox{.}}{2023}]%
        {shanahan2023role}
\bibfield{author}{\bibinfo{person}{Murray Shanahan}, \bibinfo{person}{Kyle
  McDonell}, {and} \bibinfo{person}{Laria Reynolds}.}
  \bibinfo{year}{2023}\natexlab{}.
\newblock \showarticletitle{Role play with large language models}.
\newblock \bibinfo{journal}{\emph{Nature}} \bibinfo{volume}{623},
  \bibinfo{number}{7987} (\bibinfo{year}{2023}), \bibinfo{pages}{493--498}.
\newblock


\bibitem[\protect\citeauthoryear{Shankarapani, Ramamoorthy, Movva, and
  Mukkamala}{Shankarapani et~al\mbox{.}}{2011}]%
        {shankarapani2011malware}
\bibfield{author}{\bibinfo{person}{Madhu~K Shankarapani},
  \bibinfo{person}{Subbu Ramamoorthy}, \bibinfo{person}{Ram~S Movva}, {and}
  \bibinfo{person}{Srinivas Mukkamala}.} \bibinfo{year}{2011}\natexlab{}.
\newblock \showarticletitle{Malware detection using assembly and API call
  sequences}.
\newblock \bibinfo{journal}{\emph{Journal in computer virology}}
  \bibinfo{volume}{7} (\bibinfo{year}{2011}), \bibinfo{pages}{107--119}.
\newblock


\bibitem[\protect\citeauthoryear{Shestov, Levichev, Mussabayev, and
  Maslov}{Shestov et~al\mbox{.}}{2024}]%
        {Shestov2024Finetuning}
\bibfield{author}{\bibinfo{person}{Aleksei Shestov}, \bibinfo{person}{Rodion
  Levichev}, \bibinfo{person}{Ravil Mussabayev}, {and} \bibinfo{person}{Evgeny
  Maslov}.} \bibinfo{year}{2024}\natexlab{}.
\newblock \bibinfo{title}{Finetuning Large Language Models for Vulnerability
  Detection}.
\newblock
\newblock
\urldef\tempurl%
\url{https://doi.org/10.13140/RG.2.2.26099.54560}
\showDOI{\tempurl}


\bibitem[\protect\citeauthoryear{Shi, Xiao, Wu, Zhou, Fan, and Zhang}{Shi
  et~al\mbox{.}}{2018}]%
        {shi2018pinpoint}
\bibfield{author}{\bibinfo{person}{Qingkai Shi}, \bibinfo{person}{Xiao Xiao},
  \bibinfo{person}{Rongxin Wu}, \bibinfo{person}{Jinguo Zhou},
  \bibinfo{person}{Gang Fan}, {and} \bibinfo{person}{Charles Zhang}.}
  \bibinfo{year}{2018}\natexlab{}.
\newblock \showarticletitle{Pinpoint: Fast and Precise Sparse Value Flow
  Analysis for Million Lines of Code}. In \bibinfo{booktitle}{\emph{Proceedings
  of the 39th ACM SIGPLAN Conference on Programming Language Design and
  Implementation}} (Philadelphia, PA, USA) \emph{(\bibinfo{series}{PLDI
  2018})}. \bibinfo{publisher}{Association for Computing Machinery},
  \bibinfo{address}{New York, NY, USA}, \bibinfo{pages}{693–706}.
\newblock
\showISBNx{9781450356985}
\urldef\tempurl%
\url{https://doi.org/10.1145/3192366.3192418}
\showDOI{\tempurl}


\bibitem[\protect\citeauthoryear{Song, Gao, Li, Chin, and Roychoudhury}{Song
  et~al\mbox{.}}{2024}]%
        {song2024provenfix}
\bibfield{author}{\bibinfo{person}{Yahui Song}, \bibinfo{person}{Xiang Gao},
  \bibinfo{person}{Wenhua Li}, \bibinfo{person}{Wei-Ngan Chin}, {and}
  \bibinfo{person}{Abhik Roychoudhury}.} \bibinfo{year}{2024}\natexlab{}.
\newblock \showarticletitle{ProveNFix: Temporal Property-Guided Program
  Repair}.
\newblock \bibinfo{journal}{\emph{Proceedings of the ACM on Software
  Engineering}} \bibinfo{volume}{1}, \bibinfo{number}{FSE}
  (\bibinfo{year}{2024}), \bibinfo{pages}{226--248}.
\newblock


\bibitem[\protect\citeauthoryear{Steenhoek, Gao, and Le}{Steenhoek
  et~al\mbox{.}}{2024a}]%
        {steenhoek2024dataflow}
\bibfield{author}{\bibinfo{person}{Benjamin Steenhoek},
  \bibinfo{person}{Hongyang Gao}, {and} \bibinfo{person}{Wei Le}.}
  \bibinfo{year}{2024}\natexlab{a}.
\newblock \showarticletitle{Dataflow analysis-inspired deep learning for
  efficient vulnerability detection}. In \bibinfo{booktitle}{\emph{Proceedings
  of the 46th IEEE/ACM International Conference on Software Engineering}}.
  \bibinfo{pages}{1--13}.
\newblock


\bibitem[\protect\citeauthoryear{Steenhoek, Rahman, Roy, Alam, Barr, and
  Le}{Steenhoek et~al\mbox{.}}{2024b}]%
        {DBLP:journals/corr/abs-2403-17218}
\bibfield{author}{\bibinfo{person}{Benjamin Steenhoek},
  \bibinfo{person}{Md~Mahbubur Rahman}, \bibinfo{person}{Monoshi~Kumar Roy},
  \bibinfo{person}{Mirza~Sanjida Alam}, \bibinfo{person}{Earl~T. Barr}, {and}
  \bibinfo{person}{Wei Le}.} \bibinfo{year}{2024}\natexlab{b}.
\newblock \showarticletitle{A Comprehensive Study of the Capabilities of Large
  Language Models for Vulnerability Detection}.
\newblock \bibinfo{journal}{\emph{CoRR}}  \bibinfo{volume}{abs/2403.17218}
  (\bibinfo{year}{2024}).
\newblock
\urldef\tempurl%
\url{https://doi.org/10.48550/arXiv.2403.17218}
\showURL{%
\tempurl}


\bibitem[\protect\citeauthoryear{Steenhoek, Rahman, Roy, Alam, Barr, and
  Le}{Steenhoek et~al\mbox{.}}{2024c}]%
        {steenhoek2024comprehensive}
\bibfield{author}{\bibinfo{person}{Benjamin Steenhoek},
  \bibinfo{person}{Md~Mahbubur Rahman}, \bibinfo{person}{Monoshi~Kumar Roy},
  \bibinfo{person}{Mirza~Sanjida Alam}, \bibinfo{person}{Earl~T Barr}, {and}
  \bibinfo{person}{Wei Le}.} \bibinfo{year}{2024}\natexlab{c}.
\newblock \showarticletitle{A Comprehensive Study of the Capabilities of Large
  Language Models for Vulnerability Detection}.
\newblock \bibinfo{journal}{\emph{arXiv preprint arXiv:2403.17218}}
  (\bibinfo{year}{2024}).
\newblock


\bibitem[\protect\citeauthoryear{Su and Wu}{Su and Wu}{2023}]%
        {su2023optimizing}
\bibfield{author}{\bibinfo{person}{Jingyi Su} {and} \bibinfo{person}{Yan Wu}.}
  \bibinfo{year}{2023}\natexlab{}.
\newblock \showarticletitle{Optimizing Pre-trained Language Models for
  Efficient Vulnerability Detection in Code Snippets}. In
  \bibinfo{booktitle}{\emph{2023 9th International Conference on Computer and
  Communications (ICCC)}}. IEEE, \bibinfo{pages}{2139--2143}.
\newblock


\bibitem[\protect\citeauthoryear{Sun, Liu, Iter, Zhu, and Iyyer}{Sun
  et~al\mbox{.}}{2023}]%
        {sun2023does}
\bibfield{author}{\bibinfo{person}{Simeng Sun}, \bibinfo{person}{Yang Liu},
  \bibinfo{person}{Dan Iter}, \bibinfo{person}{Chenguang Zhu}, {and}
  \bibinfo{person}{Mohit Iyyer}.} \bibinfo{year}{2023}\natexlab{}.
\newblock \showarticletitle{How does in-context learning help prompt tuning?}
\newblock \bibinfo{journal}{\emph{arXiv preprint arXiv:2302.11521}}
  (\bibinfo{year}{2023}).
\newblock


\bibitem[\protect\citeauthoryear{Tamberg and Bahsi}{Tamberg and Bahsi}{2025}]%
        {Tamberg_2025}
\bibfield{author}{\bibinfo{person}{Karl Tamberg} {and}
  \bibinfo{person}{Hayretdin Bahsi}.} \bibinfo{year}{2025}\natexlab{}.
\newblock \showarticletitle{Harnessing Large Language Models for Software
  Vulnerability Detection: A Comprehensive Benchmarking Study}.
\newblock \bibinfo{journal}{\emph{IEEE Access}}  \bibinfo{volume}{13}
  (\bibinfo{year}{2025}), \bibinfo{pages}{29698–29717}.
\newblock
\showISSN{2169-3536}
\urldef\tempurl%
\url{https://doi.org/10.1109/access.2025.3541146}
\showDOI{\tempurl}


\bibitem[\protect\citeauthoryear{Team}{Team}{2024}]%
        {team_joern_2024}
\bibfield{author}{\bibinfo{person}{T.~J. Team}.}
  \bibinfo{year}{2024}\natexlab{}.
\newblock \bibinfo{title}{Joern}.
\newblock \bibinfo{howpublished}{Online}.
\newblock
\newblock
\shownote{Available: \url{https://joern.io}.}


\bibitem[\protect\citeauthoryear{Touvron, Lavril, Izacard, Martinet, Lachaux,
  Lacroix, Rozi{\`e}re, Goyal, Hambro, Azhar, et~al\mbox{.}}{Touvron
  et~al\mbox{.}}{2023}]%
        {touvron2023llama}
\bibfield{author}{\bibinfo{person}{Hugo Touvron}, \bibinfo{person}{Thibaut
  Lavril}, \bibinfo{person}{Gautier Izacard}, \bibinfo{person}{Xavier
  Martinet}, \bibinfo{person}{Marie-Anne Lachaux},
  \bibinfo{person}{Timoth{\'e}e Lacroix}, \bibinfo{person}{Baptiste
  Rozi{\`e}re}, \bibinfo{person}{Naman Goyal}, \bibinfo{person}{Eric Hambro},
  \bibinfo{person}{Faisal Azhar}, {et~al\mbox{.}}}
  \bibinfo{year}{2023}\natexlab{}.
\newblock \showarticletitle{Llama: Open and efficient foundation language
  models}.
\newblock \bibinfo{journal}{\emph{arXiv preprint arXiv:2302.13971}}
  (\bibinfo{year}{2023}).
\newblock


\bibitem[\protect\citeauthoryear{Wang, Li, Li, Xiong, Li, Yan, and Jin}{Wang
  et~al\mbox{.}}{2024}]%
        {wang2024m2cvdenhancingvulnerabilitysemantic}
\bibfield{author}{\bibinfo{person}{Ziliang Wang}, \bibinfo{person}{Ge Li},
  \bibinfo{person}{Jia Li}, \bibinfo{person}{Yingfei Xiong},
  \bibinfo{person}{Jia Li}, \bibinfo{person}{Meng Yan}, {and}
  \bibinfo{person}{Zhi Jin}.} \bibinfo{year}{2024}\natexlab{}.
\newblock \bibinfo{title}{M2CVD: Enhancing Vulnerability Semantic through
  Multi-Model Collaboration for Code Vulnerability Detection}.
\newblock
\newblock
\showeprint[arxiv]{2406.05940}~[cs.SE]
\urldef\tempurl%
\url{https://arxiv.org/abs/2406.05940}
\showURL{%
\tempurl}


\bibitem[\protect\citeauthoryear{Wei, Wang, Schuurmans, Bosma, Xia, Chi, Le,
  Zhou, et~al\mbox{.}}{Wei et~al\mbox{.}}{2022}]%
        {wei2022chain}
\bibfield{author}{\bibinfo{person}{Jason Wei}, \bibinfo{person}{Xuezhi Wang},
  \bibinfo{person}{Dale Schuurmans}, \bibinfo{person}{Maarten Bosma},
  \bibinfo{person}{Fei Xia}, \bibinfo{person}{Ed Chi}, \bibinfo{person}{Quoc~V
  Le}, \bibinfo{person}{Denny Zhou}, {et~al\mbox{.}}}
  \bibinfo{year}{2022}\natexlab{}.
\newblock \showarticletitle{Chain-of-thought prompting elicits reasoning in
  large language models}.
\newblock \bibinfo{journal}{\emph{Advances in neural information processing
  systems}}  \bibinfo{volume}{35} (\bibinfo{year}{2022}),
  \bibinfo{pages}{24824--24837}.
\newblock


\bibitem[\protect\citeauthoryear{Wen, Wang, Chen, Hu, Lo, and Gao}{Wen
  et~al\mbox{.}}{2024}]%
        {wen2024vuleval}
\bibfield{author}{\bibinfo{person}{Xin-Cheng Wen}, \bibinfo{person}{Xinchen
  Wang}, \bibinfo{person}{Yujia Chen}, \bibinfo{person}{Ruida Hu},
  \bibinfo{person}{David Lo}, {and} \bibinfo{person}{Cuiyun Gao}.}
  \bibinfo{year}{2024}\natexlab{}.
\newblock \showarticletitle{Vuleval: Towards repository-level evaluation of
  software vulnerability detection}.
\newblock \bibinfo{journal}{\emph{arXiv preprint arXiv:2404.15596}}
  (\bibinfo{year}{2024}).
\newblock


\bibitem[\protect\citeauthoryear{Wu and Zou}{Wu and Zou}{2022}]%
        {wu2022code}
\bibfield{author}{\bibinfo{person}{Bolun Wu} {and} \bibinfo{person}{Futai
  Zou}.} \bibinfo{year}{2022}\natexlab{}.
\newblock \showarticletitle{Code vulnerability detection based on deep sequence
  and graph models: A survey}.
\newblock \bibinfo{journal}{\emph{Security and Communication Networks}}
  \bibinfo{volume}{2022}, \bibinfo{number}{1} (\bibinfo{year}{2022}),
  \bibinfo{pages}{1176898}.
\newblock


\bibitem[\protect\citeauthoryear{Yamaguchi, Golde, Arp, and Rieck}{Yamaguchi
  et~al\mbox{.}}{2014}]%
        {yamaguchi2014modeling}
\bibfield{author}{\bibinfo{person}{Fabian Yamaguchi}, \bibinfo{person}{Nico
  Golde}, \bibinfo{person}{Daniel Arp}, {and} \bibinfo{person}{Konrad Rieck}.}
  \bibinfo{year}{2014}\natexlab{}.
\newblock \showarticletitle{Modeling and discovering vulnerabilities with code
  property graphs}. In \bibinfo{booktitle}{\emph{2014 IEEE symposium on
  security and privacy}}. IEEE, \bibinfo{pages}{590--604}.
\newblock


\bibitem[\protect\citeauthoryear{Yang, Zhou, Mao, Xu, Yang, Zhang, Shen, and
  Zhang}{Yang et~al\mbox{.}}{2024}]%
        {yang2024dlap}
\bibfield{author}{\bibinfo{person}{Yanjing Yang}, \bibinfo{person}{Xin Zhou},
  \bibinfo{person}{Runfeng Mao}, \bibinfo{person}{Jinwei Xu},
  \bibinfo{person}{Lanxin Yang}, \bibinfo{person}{Yu Zhang},
  \bibinfo{person}{Haifeng Shen}, {and} \bibinfo{person}{He Zhang}.}
  \bibinfo{year}{2024}\natexlab{}.
\newblock \showarticletitle{Dlap: A deep learning augmented large language
  model prompting framework for software vulnerability detection}.
\newblock \bibinfo{journal}{\emph{Journal of Systems and Software}}
  (\bibinfo{year}{2024}), \bibinfo{pages}{112234}.
\newblock


\bibitem[\protect\citeauthoryear{Yang, Zhou, Mao, Xu, Yang, Zhang, Shen, and
  Zhang}{Yang et~al\mbox{.}}{2025}]%
        {10.1016/j.jss.2024.112234}
\bibfield{author}{\bibinfo{person}{Yanjing Yang}, \bibinfo{person}{Xin Zhou},
  \bibinfo{person}{Runfeng Mao}, \bibinfo{person}{Jinwei Xu},
  \bibinfo{person}{Lanxin Yang}, \bibinfo{person}{Yu Zhang},
  \bibinfo{person}{Haifeng Shen}, {and} \bibinfo{person}{He Zhang}.}
  \bibinfo{year}{2025}\natexlab{}.
\newblock \showarticletitle{DLAP: A Deep Learning Augmented Large Language
  Model Prompting framework for software vulnerability detection}.
\newblock \bibinfo{journal}{\emph{J. Syst. Softw.}} \bibinfo{volume}{219},
  \bibinfo{number}{C} (\bibinfo{date}{Jan.} \bibinfo{year}{2025}),
  \bibinfo{numpages}{15}~pages.
\newblock
\showISSN{0164-1212}
\urldef\tempurl%
\url{https://doi.org/10.1016/j.jss.2024.112234}
\showDOI{\tempurl}


\bibitem[\protect\citeauthoryear{Zhang, Liu, Zeng, Yang, Li, and Li}{Zhang
  et~al\mbox{.}}{2024b}]%
        {10.1145/3639478.3643065}
\bibfield{author}{\bibinfo{person}{Chenyuan Zhang}, \bibinfo{person}{Hao Liu},
  \bibinfo{person}{Jiutian Zeng}, \bibinfo{person}{Kejing Yang},
  \bibinfo{person}{Yuhong Li}, {and} \bibinfo{person}{Hui Li}.}
  \bibinfo{year}{2024}\natexlab{b}.
\newblock \showarticletitle{Prompt-Enhanced Software Vulnerability Detection
  Using ChatGPT}. In \bibinfo{booktitle}{\emph{Proceedings of the 2024 IEEE/ACM
  46th International Conference on Software Engineering: Companion
  Proceedings}} (Lisbon, Portugal) \emph{(\bibinfo{series}{ICSE-Companion
  '24})}. \bibinfo{publisher}{Association for Computing Machinery},
  \bibinfo{address}{New York, NY, USA}, \bibinfo{pages}{276–277}.
\newblock
\showISBNx{9798400705021}
\urldef\tempurl%
\url{https://doi.org/10.1145/3639478.3643065}
\showDOI{\tempurl}


\bibitem[\protect\citeauthoryear{Zhang, Liu, Zeng, Yang, Li, and Li}{Zhang
  et~al\mbox{.}}{2024c}]%
        {zhang2024prompt}
\bibfield{author}{\bibinfo{person}{Chenyuan Zhang}, \bibinfo{person}{Hao Liu},
  \bibinfo{person}{Jiutian Zeng}, \bibinfo{person}{Kejing Yang},
  \bibinfo{person}{Yuhong Li}, {and} \bibinfo{person}{Hui Li}.}
  \bibinfo{year}{2024}\natexlab{c}.
\newblock \showarticletitle{Prompt-enhanced software vulnerability detection
  using chatgpt}. In \bibinfo{booktitle}{\emph{Proceedings of the 2024 IEEE/ACM
  46th International Conference on Software Engineering: Companion
  Proceedings}}. \bibinfo{pages}{276--277}.
\newblock


\bibitem[\protect\citeauthoryear{Zhang, Ma, Hu, Liu, Xie, Le~Traon, and
  Liu}{Zhang et~al\mbox{.}}{2023a}]%
        {zhang-etal-2023-black}
\bibfield{author}{\bibinfo{person}{Jie Zhang}, \bibinfo{person}{Wei Ma},
  \bibinfo{person}{Qiang Hu}, \bibinfo{person}{Shangqing Liu},
  \bibinfo{person}{Xiaofei Xie}, \bibinfo{person}{Yves Le~Traon}, {and}
  \bibinfo{person}{Yang Liu}.} \bibinfo{year}{2023}\natexlab{a}.
\newblock \showarticletitle{A Black-Box Attack on Code Models via
  Representation Nearest Neighbor Search}. In
  \bibinfo{booktitle}{\emph{Findings of the Association for Computational
  Linguistics: EMNLP 2023}}, \bibfield{editor}{\bibinfo{person}{Houda Bouamor},
  \bibinfo{person}{Juan Pino}, {and} \bibinfo{person}{Kalika Bali}} (Eds.).
  \bibinfo{publisher}{Association for Computational Linguistics},
  \bibinfo{address}{Singapore}, \bibinfo{pages}{9706--9716}.
\newblock
\urldef\tempurl%
\url{https://doi.org/10.18653/v1/2023.findings-emnlp.649}
\showDOI{\tempurl}


\bibitem[\protect\citeauthoryear{Zhang, Hu, and Chen}{Zhang
  et~al\mbox{.}}{2024a}]%
        {zhang2024context}
\bibfield{author}{\bibinfo{person}{Yulin Zhang}, \bibinfo{person}{Yong Hu},
  {and} \bibinfo{person}{Xiao Chen}.} \bibinfo{year}{2024}\natexlab{a}.
\newblock \showarticletitle{Context and Multi-Features-Based Vulnerability
  Detection: A Vulnerability Detection Frame Based on Context Slicing and
  Multi-Features}.
\newblock \bibinfo{journal}{\emph{Sensors}} \bibinfo{volume}{24},
  \bibinfo{number}{5} (\bibinfo{year}{2024}), \bibinfo{pages}{1351}.
\newblock


\bibitem[\protect\citeauthoryear{Zhang, Zhu, Yang, Wen, and Jin}{Zhang
  et~al\mbox{.}}{2023b}]%
        {Zhang2023Compare}
\bibfield{author}{\bibinfo{person}{Yuting Zhang}, \bibinfo{person}{Jiahao Zhu},
  \bibinfo{person}{Yixin Yang}, \bibinfo{person}{Ming Wen}, {and}
  \bibinfo{person}{Hai Jin}.} \bibinfo{year}{2023}\natexlab{b}.
\newblock \showarticletitle{Comparing the Performance of Different Code
  Representations for Learning-based Vulnerability Detection}. In
  \bibinfo{booktitle}{\emph{Proceedings of the 14th Asia-Pacific Symposium on
  Internetware}} (Hangzhou, China) \emph{(\bibinfo{series}{Internetware '23})}.
  \bibinfo{publisher}{Association for Computing Machinery},
  \bibinfo{address}{New York, NY, USA}, \bibinfo{pages}{174–184}.
\newblock
\showISBNx{9798400708947}
\urldef\tempurl%
\url{https://doi.org/10.1145/3609437.3609464}
\showDOI{\tempurl}


\bibitem[\protect\citeauthoryear{Zhou, Chen, Liu, Ackah-Arthur, Chen, Zhang,
  and Zeng}{Zhou et~al\mbox{.}}{2019a}]%
        {zhou2019method}
\bibfield{author}{\bibinfo{person}{Minmin Zhou}, \bibinfo{person}{Jinfu Chen},
  \bibinfo{person}{Yisong Liu}, \bibinfo{person}{Hilary Ackah-Arthur},
  \bibinfo{person}{Shujie Chen}, \bibinfo{person}{Qingchen Zhang}, {and}
  \bibinfo{person}{Zhifeng Zeng}.} \bibinfo{year}{2019}\natexlab{a}.
\newblock \showarticletitle{A method for software vulnerability detection based
  on improved control flow graph}.
\newblock \bibinfo{journal}{\emph{Wuhan University Journal of Natural
  Sciences}} \bibinfo{volume}{24}, \bibinfo{number}{2} (\bibinfo{year}{2019}),
  \bibinfo{pages}{149--160}.
\newblock


\bibitem[\protect\citeauthoryear{Zhou, Tran, Le-Cong, Zhang, Irsan, Sumarlin,
  Le, and Lo}{Zhou et~al\mbox{.}}{2024a}]%
        {zhou2024comparison}
\bibfield{author}{\bibinfo{person}{Xin Zhou}, \bibinfo{person}{Duc-Manh Tran},
  \bibinfo{person}{Thanh Le-Cong}, \bibinfo{person}{Ting Zhang},
  \bibinfo{person}{Ivana~Clairine Irsan}, \bibinfo{person}{Joshua Sumarlin},
  \bibinfo{person}{Bach Le}, {and} \bibinfo{person}{David Lo}.}
  \bibinfo{year}{2024}\natexlab{a}.
\newblock \showarticletitle{Comparison of static application security testing
  tools and large language models for repo-level vulnerability detection}.
\newblock \bibinfo{journal}{\emph{arXiv preprint arXiv:2407.16235}}
  (\bibinfo{year}{2024}).
\newblock


\bibitem[\protect\citeauthoryear{Zhou, Zhang, and Lo}{Zhou
  et~al\mbox{.}}{2024b}]%
        {10.1145/3639476.3639762}
\bibfield{author}{\bibinfo{person}{Xin Zhou}, \bibinfo{person}{Ting Zhang},
  {and} \bibinfo{person}{David Lo}.} \bibinfo{year}{2024}\natexlab{b}.
\newblock \showarticletitle{Large Language Model for Vulnerability Detection:
  Emerging Results and Future Directions}. In
  \bibinfo{booktitle}{\emph{Proceedings of the 2024 ACM/IEEE 44th International
  Conference on Software Engineering: New Ideas and Emerging Results}} (Lisbon,
  Portugal) \emph{(\bibinfo{series}{ICSE-NIER'24})}.
  \bibinfo{publisher}{Association for Computing Machinery},
  \bibinfo{address}{New York, NY, USA}, \bibinfo{pages}{47–51}.
\newblock
\showISBNx{9798400705007}
\urldef\tempurl%
\url{https://doi.org/10.1145/3639476.3639762}
\showDOI{\tempurl}


\bibitem[\protect\citeauthoryear{Zhou, Zhang, and Lo}{Zhou
  et~al\mbox{.}}{2024c}]%
        {zhou2024large}
\bibfield{author}{\bibinfo{person}{Xin Zhou}, \bibinfo{person}{Ting Zhang},
  {and} \bibinfo{person}{David Lo}.} \bibinfo{year}{2024}\natexlab{c}.
\newblock \showarticletitle{Large language model for vulnerability detection:
  Emerging results and future directions}. In
  \bibinfo{booktitle}{\emph{Proceedings of the 2024 ACM/IEEE 44th International
  Conference on Software Engineering: New Ideas and Emerging Results}}.
  \bibinfo{pages}{47--51}.
\newblock


\bibitem[\protect\citeauthoryear{Zhou, Liu, Siow, Du, and Liu}{Zhou
  et~al\mbox{.}}{2019b}]%
        {zhou2019devign}
\bibfield{author}{\bibinfo{person}{Yaqin Zhou}, \bibinfo{person}{Shangqing
  Liu}, \bibinfo{person}{Jingkai Siow}, \bibinfo{person}{Xiaoning Du}, {and}
  \bibinfo{person}{Yang Liu}.} \bibinfo{year}{2019}\natexlab{b}.
\newblock \showarticletitle{Devign: Effective vulnerability identification by
  learning comprehensive program semantics via graph neural networks}.
\newblock \bibinfo{journal}{\emph{Advances in neural information processing
  systems}}  \bibinfo{volume}{32} (\bibinfo{year}{2019}).
\newblock


\bibitem[\protect\citeauthoryear{Zimmermann, Premraj, and Zeller}{Zimmermann
  et~al\mbox{.}}{2007}]%
        {zimmermann2007predicting}
\bibfield{author}{\bibinfo{person}{Thomas Zimmermann}, \bibinfo{person}{Rahul
  Premraj}, {and} \bibinfo{person}{Andreas Zeller}.}
  \bibinfo{year}{2007}\natexlab{}.
\newblock \showarticletitle{Predicting defects for eclipse}. In
  \bibinfo{booktitle}{\emph{Third International Workshop on Predictor Models in
  Software Engineering (PROMISE'07: ICSE Workshops 2007)}}. IEEE,
  \bibinfo{pages}{9--9}.
\newblock


\end{thebibliography}



\end{document}